\documentclass{article}
\usepackage[utf8]{inputenc}
\usepackage{hyperref}
\pdfoutput=1
\usepackage[cm]{fullpage}
\usepackage{longtable}
\usepackage{ifthen}
\title{Stop the Open Data Bus, We Want to Get Off}
\author{Dr. Chris Culnane, A/Prof. Benjamin I. P. Rubinstein, A/Prof. Vanessa Teague \\ The University of Melbourne, Australia \\
\texttt{[cculnane, brubinstein, vjteague]@unimelb.edu.au}}
\date{August 15, 2019}

\usepackage[frozencache=true]{minted}
\usepackage[disable]{todonotes}

\newcommand{\commentOut}[1]{}

\usepackage{graphicx}
\usepackage{caption}
\usepackage{subcaption}
\usepackage{comment}

\newcounter{RELEASETYPE}
\ifthenelse{\value{RELEASETYPE}=0}{
\includecomment{fulldetails}
\includecomment{partialdetails}
\excludecomment{safecontent}
}{}
\ifthenelse{\value{RELEASETYPE}=1}{
\excludecomment{fulldetails}
\includecomment{partialdetails}
\excludecomment{safecontent}
}{}
\ifthenelse{\value{RELEASETYPE}=2}{
\excludecomment{fulldetails}
\excludecomment{partialdetails}
\includecomment{safecontent}
}{}

\newcommand{\release}[3][]{\ifthenelse{\equal{#1}{}}{\ifthenelse{\value{RELEASETYPE}=2}{#2}{#3}}{\ifthenelse{\value{RELEASETYPE}=#1 \or \value{RELEASETYPE}>#1}{#2}{#3}}}

\begin{document}

\maketitle

\section{Introduction}

The subject of this report is the re-identification of individuals in the Myki public transport dataset released as part of the Melbourne Datathon 2018. We demonstrate the ease with which we were able to re-identify ourselves, our co-travellers, and complete strangers; our analysis  raises concerns about the nature and granularity of the data released, in particular the ability to identify vulnerable or sensitive groups.

\paragraph{Report organisation.} We provide a short introduction outlining how the data was released and how we obtained a copy of it. The first part of our analysis focuses on whether we could re-identify people within the data release, starting with ourselves. The second part of our analysis looks at the broader issue of identifiability in transport data.  In Section~\ref{sec:reid} we explain how we re-identified ourselves. Section~\ref{sec:cotravel} covers co-traveller analysis, and how we could combine that with our own re-identification to re-identify a further individual as a result of a single co-travel event.
Section~\ref{sec:sensitive} highlights concerns regarding the ability to identity vulnerable groups of people. Section~\ref{subsec:MPRe-Id} shows how an individual can be re-identified from additional public information (in this case, his tweets).  Section~\ref{sec:uniqueness} is a general analysis of uniqueness in the dataset.  Recommendations on better technology and process are offered in Section~\ref{sec:better}.  We discuss the broader context of Australian privacy laws and data sharing practices  in Section~\ref{sec:discussion}.  Finally, Section~\ref{sec:conclusion}  summarises our findings and concerns.

\subsection{Related Work on Re-identification}
De-identification of detailed unit-record-level data does not work, at least not without substantially altering the data to the point where its general value is significantly reduced~\cite{ovic2018}.  This has been demonstrated in numerous studies across domains including health data~\cite{sweeney2000uniqueness, culnane2017health}, film preferences~\cite{narayanan2008robust}, social connections~\cite{narayanan2009anonymizing}, location~\cite{de2013unique}, spending patterns~\cite{de2015unique}, search query logs~\cite{Barbaro2006aol}, and many others.

A common theme is that a remarkably small number of distinct points of information are necessary to make an individual
unique---whenever one person's information is linked together into a detailed record of their events, a few known
events are usually enough to identify them.  De~Montjoye \emph{et al.}~\cite{de2013unique} showed that 80\% of telephone
users were unique based on  3 points of time and location, even when neither times nor places were very precisely given.

An analysis of a now infamous 2014 Freedom of Information release of New York taxi data revealed that even passenger-centric transport data could be identifying~\cite{Pandurangan2014taxi}.

Another common theme is that most putative anonymisation approaches do not achieve any security property in particular, but rather rely on apparent complexity to prevent reversal via re-identification.   This generally does not work.
For example, the U.S. Census Bureau recently performed a landmark reconstruction attack on the 2010 U.S. Census release, achieving a high level of effectiveness~\cite{garfinkel2019understanding}: confirmed reconstruction of PII for 17\% of the U.S. population.
Privacy can be properly protected only by using technical privacy frameworks that formally guarantee a specific security property against a well-defined threat model of adversary capabilities and information.  All products from the 2020 U.S. Census will be released under the differential privacy framework~\cite{abowd2018us}.

\subsection{Implications of This Work}
 It is rare for researchers to have access to such a large and detailed transport dataset. By analysing this release we can learn about the broader issues of privacy in smart ticketing schemes, particularly those that are facilitating the inclusion of commercial card providers into their ecosystems.  Myki, for example, now offers a Google Pay option.\footnote{\url{https://www.ptv.vic.gov.au/tickets/myki/mobile-myki/}}  Thus a great deal of detailed information is already being provided to third parties, regardless of whether it is posted on the web.

 This work highlights how a large number of passengers could be re-identified in the 2018 Myki data release, with detailed discussion of specific people.
  The implications of re-identification are potentially serious: ex-partners, one-time acquaintances, or other parties can determine places of home, work, times of travel, co-travelling patterns---presenting risk to vulnerable groups in particular.

\section{The Myki Data Release}
In 2018 the Victorian Government released a large passenger centric transport dataset to a data science competition---the 2018 Melbourne Datathon. Access to the data was unrestricted, with a URL provided on the datathon's website to download the complete dataset from an Amazon S3 Bucket. Over 190 teams continued to analyse the data through the 2 month competition period. The data consisted of touch on and touch off events for the Myki smart card ticketing system used throughout the state of Victoria, Australia. With such data, contestants would be able to apply retrospective analyses on an entire public transport system, explore suitability of predictive models, etc.

The Myki ticketing system is used across Victorian public transport: on trains, buses and trams. The dataset was a longitudinal dataset, consisting of touch on and touch off events from Week 27 in 2015 through to Week 26 in 2018. Each event contained a card identifier (cardId; not the actual card number), the card type, the time of the touch on or off, and various location information, for example a stop ID or route ID, along with other fields which we omit here for brevity. Events could be indexed by the cardId and as such, all the events associated with a single card could be retrieved. There are a total of 15,184,336 cards in the dataset---more than twice the 2018 population of
Victoria.  It appears that all touch on and off events for metropolitan trains and trams have been included, though other forms of transport such as intercity
trains and some buses are absent. In total there are nearly 2 billion
touch on and off events in the dataset.

No information was provided as to the de-identification that was performed on the dataset. Our analysis indicates that little to no de-identification took place on the bulk of the data, as will become evident in Section~\ref{sec:data_analysis}. The exception is the cardId, which appears to have been mapped in some way from the Myki Card Number. The exact mapping has not been discovered, although concerns remain as to its security effectiveness.

\subsection{Card ID De-identification}
No details were publicly given about how the cardId field was generated, but it is not the Myki Card Number, nor does it appear to be a random number. Our analysis of the cardId and its distribution indicated some form of mapping was applied. The cardId is 8 digits, but the range of values in the dataset is only from 1 to 24451922. Such a low maximum value is in itself unusual, given that there are over 15 million cards in the dataset, and the maximum Myki Card Number far exceeds the maximum value in the dataset. Furthermore, the cardIds are not uniformly distributed throughout the available space, particularly towards the lower end of the range. For example, there are no cardIds between 2 and 15746. It could be that the cardId of 1 is an error, but that would not explain why the first fifteen thousand possible IDs are not used. This unexpectedly large gap between cardIds re-occurs, as shown in Table~\ref{tab:cardgaps}.
\begin{table}[ht]
\centering
\begin{tabular}{|c|c|c|}
 \hline
 \emph{Start of Gap} & \emph{End of Gap} & \emph{Gap Size}  \\
 \hline
 154449 & 173211 & 18764  \\
\hline
 173338 & 180637 & 7297  \\
\hline
 191920 & 356913 & 164995  \\
\hline
\end{tabular}
\caption{Non-Uniform Distribution of cardIds.}
\label{tab:cardgaps}
\end{table}

There are a number of such gaps in the first 60,000 cardIds. Such gaps are not consistent with a random allocation of IDs, and indicates some form of structure being applied to the cardId. There also appears to be an element of time associated with the ID. For example, the last 180,000 cardIds all have a first touch on event after 1st May 2018. Whilst concerns remain about the security of the mapping to cardId, it was not the focus of our analysis.

\section{Data Analysis}\label{sec:data_analysis}
In this section we detail the ways in which private or public information can be used to re-identify specific individuals in the data.
\subsection{Re-identifying Ourselves}\label{sec:reid}
It was a straightforward task to re-identify two of the co-authors. Both of us have registered our Myki cards, which provides online access to the last 6 months of trip data, with precise, down to the second, touch on and touch off event information. By searching for a card with a matching touch on time shown in Ben's trip data we find 48 possible matches. By searching for a card with just two of Ben's touch on times we get a single unique match. Both touch on times that were searched for were what would be considered busy commuter times (between 8am and 9am, and between 5pm and 6pm).\footnote{We have not provided exact times to mitigate re-identification by others.} The same was true for Chris's card, an initial query for a morning trip between 7am and 8am returned 59 possible matches, adding just one more trip, this time an evening trip between 7pm and 8pm, revealed an exact match.

To further confirm the cards found were our own, we checked the first times they appeared in the dataset, which in both cases corresponded to when the cards were purchased/activated. We also looked at a sample of trips taken, which were consistent with our respective travel patterns.

This demonstrates that anyone who has registered their Myki card is able to  easily re-identify themselves within this dataset. That in itself may not be considered damaging, since in order to perform the re-identification one would require access to the equivalent data via the Myki account website. However, as we shall demonstrate, the ability to re-identify oneself can be leveraged to re-identify others, including those who have not registered their cards.

\subsection{Co-traveller Analysis}\label{sec:cotravel}
In order to leverage our ability to  and identify our own cards we performed a novel co-traveller analysis to determine if we could find others who we had co-travelled with. We define a co-traveller as anyone who touches on at the same stop and time, $\pm$ 5 seconds. When we performed the analysis for the period from 2017 onwards (approximately 18 months) we produced a list of co-travellers for each card that contained 2106 rows for Chris, and 8591 co-travellers for Ben. The difference is a reflection of the different means of public transport used, with Chris's predominantly being trams, whilst Ben's combines trams and trains. However, in both cases the number of repeat co-travelling cases was relatively small. Chris had 38 co-travelling cards with an occurrence of 2 or more, whilst Ben had 363 co-travelling cards with an occurrence of 2 or more. To further strengthen our confidence in our re-identification of ourselves, we could see that we appeared in each other's co-travellers lists. In Chris's case, Ben was the most frequent co-traveller with 7 occurrences in the period, and Chris was the 4th most regular co-traveller for Ben. We further analysed the co-travelling occurrences between ourselves, and were able to cross-match them to events/meetings we had attended together in Melbourne's central business district.

Co-travelling occurrences could reveal related individuals, for example children or partners. For example, the third most regular co-traveller for Ben is a Child concession card, strongly indicating a family connection.

\subsubsection{Co-traveller Re-identification}
It is not just regular co-travellers that can be re-identified, a single co-travelling event can be sufficient to re-identify an individual. In this instance we looked for an individual, Peter Tonoli, who Chris had co-travelled with on one known occasion. Peter kindly consented to be named within this report. Using minimal background information, namely that Peter is a commuter, and a broad idea of where he lives, we were able to successfully re-identify him. Firstly we will provide some background as to why we can be certain of the date on which Peter was a co-traveller with Chris. The background is that a group of us had attended an evening seminar, followed by a social gathering afterwards. Chris and Peter left at the same time and boarded the same tram. This is the only occasion in which they have knowingly travelled together.\footnote{We were able to find a second occurrence when they appear to have been on the same tram, but were unaware of it.} Using our own prior knowledge we could look up the date of the seminar in our calendars. Using the ability to re-identify Chris's card, we were able to look for all single co-travel occurrences on the date of the seminar, and the corresponding times for the card touch on events. This revealed 4 possible matches. From those 4 matches, two were concession cards, which could be immediately excluded, one was a card that had only a handful of events---which would not be consistent with a commuter like Peter. This left a single possible match. To gain further assurance it was an accurate re-identification, we checked that there was regular travel to and from the station near where Peter lives, and checked with Peter the date his card expired, and therefore ceased to record any further events in the dataset. All such checks were correct, giving us a high degree of confidence in the re-identification.

This type of re-identification is particularly concerning, since it allows an individual to leverage the ease of re-identifying themselves to re-identify others, and from potentially only a single co-travel event. This presents a risk for anyone who has co-travelled with someone in the past, for example, an ex-partner, a co-worker, or even just someone they went on a single date with. Due to the large amount of data provided, \emph{i.e.,} all touch on and off events, it could allow a malicious party to determine where someone lived, worked, or socialised---and when they visit these places and for how long. This would be a particular problem for those seeking to escape an abusive partner, or those who have been subject to unwanted attention. It would not be sufficient to merely get a new card today, since the dataset will reveal the location information up until the middle of 2018.

\subsection{Myki-Specific Analysis of Sensitive Types}\label{sec:sensitive}
Up until this point our analysis would be broadly applicable across different smart ticketing systems. Most such systems will record touch on and off events along with the corresponding dates and times. We conducted further analysis of specific concerns with the Myki release. We did not seek to re-identify individuals in these instances due to the sensitivity of some of the attributes, but the fact such information was released raises additional concerns about the potential harm of the release.

The dataset includes the card type for each trip. Unfortunately, the card type can reveal sensitive attributes. There are a total of 74 different card types defined in the dataset. Whilst most will not be particularly identifying, for example, 0 represents the Default Full Fare, and 65 represents the Commuter Club. They can still assist in re-identification, for example, checking that Ben's card, which is a Commuter Club card, has a type of 65.

What is far more concerning is the card types that reveal information about the holder, for example, potentially sensitive jobs. We found 371 type 46 cards, which is Federal Police Travel Pass, and 1232 type 48 cards which is a Transit Police Travel Pass. We also found 8 type 50 cards which represents a Federal Parliamentarian Travel Pass, and 424 type 51 cards, representing a State Parliamentarian Travel Pass.

Holders of those cards may not want their travel movements to be known to others, since it could reveal broad information about where those individuals live, and their patterns of travel. Such information could present a security risk to those individuals. Furthermore, if those individuals use their cards outside of going to and from work, and they co-travel with their family, they may reveal information about the movements and travel patterns of their family members, further increasing the security risk.

Furthermore, the granularity of card types for children is also of concern, including card types for children 5 years to 18 years old, of which there were 697,219 instances, Primary Student Concession\footnote{Cards issued to primary school children, who will be aged 6 to 12} cards, of which there were 609, and Secondary Student Concession\footnote{Cards issued to secondary school children, who will be aged 12 to 16} cards, of which there were 178,842 instances. Since the holders of those cards are likely to have a regular travel pattern, there is a risk that information such as when and where a child travels, and even possibly whether a child is travelling unaccompanied, could be discerned. We have not performed such analysis, but are concerned that there appear to be no barriers preventing others from doing so.

\subsection{Re-identifying Strangers by Linking with Public Data} \label{subsec:MPRe-Id}
In this section we show that it is possible to identify a stranger from public information about their travel patterns.  In this case, our data source is Twitter, but there are numerous other possible sources.

There are 218 Melbourne metropolitan train stations and 424 State Parliamentarian Travel Passes, so it is not surprising that some (mostly outer-urban) train stations are visited by only a very small number of State Parliamentarians through the 3-year data range.  For most of these, it is probably quite obvious which Parliamentarian(s) visit there.  We illustrate one example here, though there are probably many more.

Anthony Carbines MP is the Member of the Victorian Legislative Assembly for the district of Ivanhoe, with an electorate office a few metres from Rosanna train station.  He tweets frequently about his experiences on Melbourne's train services.  We'd like to thank Mr Carbines for agreeing to have his data used as a case study in this report.

When we search the database for State Parliamentarian (type 51) cards that have visited Rosanna train station, there are only two.
The first of these has visited Rosanna only once; the second (card ID 11891903) commutes frequently from Rosanna into the city and back again.  This seems likely to be Mr Carbines.

To test the hypothesis that (ID 11891903) is Mr Carbines, we searched Twitter for his tweets about train travel, noting information about when and where he used public transport.  
\begin{partialdetails}
Three results are shown in Figure~\ref{fig:TweetsInRange}, with a note about corresponding activity on (card ID 11891903). Another 15 are contained in the Appendix.
\end{partialdetails}
All 18 match with a record in the dataset, most with a tap-on or tap-off time within ten minutes of the tweet.  It is overwhelmingly unlikely that this is a coincidental resemblance to a different person.\footnote{The only remotely plausible possible match would be a staffer who also travelled on a type-51 card and accompanied Mr Carbines everywhere.  However, this seems unlikely to produce such a perfect match on every single tweet, and does not explain why there is only one type-51 match for the Rosanna records, not two or more.} There were no tweets without a matching record in the dataset, except those mentioning buses or Geelong (which we suspect are not all in the dataset).

The tweets can be found by simply searching Twitter : \url{https://twitter.com/search?q=\%40ACarbinesMP\%20train&src=typd}
A number of other tweets mention Rosanna station or the Rosanna line, but are out of the date range of the dataset.

\begin{partialdetails}

\begin{figure}
\begin{longtable}{p{2cm}p{12cm}}
{\bf Date/Time} &  {\bf Tweet/event in data/Notes} \\
16 Mar 2016  11:43am & ``New digs in Taylors Road St Albans for local MP @NatalieSuleyman the train takes me over Main \& Furlong Rds action!'' \\
     &    \texttt{onDate:2016-03-16 11:29:06. onStopId:2000 Keilor Plains} \\
     &  \texttt{offDate:2016-03-16 12:07:37.
      offStopId:64403 City} \\
      &  Consistent record (14 mins time gap) for coming from St Albans to the city. Stop  20001 is Keilor Plains Railway Station (St Albans), the closest station to her electorate office.\\
           & \\
3 May 2016   6:57am & ``No busy AM train to the city...Greensie bound to meet local school principals with @VickiWardMP @ColinBrooksMP'' \\
   &  \texttt{onDate:2016-05-03 06:53:22.  onStopId:{\bf 19936} Rosanna} \\
   & \texttt{offDate:2016-05-03 07:12:21. offStopId:19985 Greensborough }\\
   &  Perfectly matching (4 mins).
   %19985 is Greensborough station.
   \\
        & \\
4 May 2018   & ``Rosanna station. 5.24AM Touching on.''  \\
5:34am    &  \texttt{onDate:2018-05-04 05:11:18. onStopId:{\bf 19936} Rosanna }\\
   & \texttt{offDate:2018-05-04 05:40:54. offStopId : 19929 Dennis.} \\
   & \texttt{onDate:2018-05-04 05:41:52. onStopId : 19929 Dennis.} \\
   & \texttt{offDate:2018-05-04 06:07:17. offStopId:19983} \\
    &  Matching times (13 mins).  Not clear why there are 2 records---seems to have tapped off at 19929 (Dennis) and immediately tapped on again.\\[0.5em]
\end{longtable}
\caption{Three example tweets mentioning different train stations.}
\label{fig:TweetsInRange}
\end{figure}

\end{partialdetails}

There is no other type-51 card at Rosanna station on 4 May 2018, so this single data point is enough to identify Mr Carbines's card and be highly confident of a correct match.

\subsubsection{Examining Re-identification Without Special Information}
We are very confident of having correctly identified Mr Carbines from 18 matching events, but it is interesting to consider what could have been inferred from  the three tweets in Figure~\ref{fig:TweetsInRange} alone.

Although we used precise times and the fact that the card was a state MP (type 51), this  wasn't necessary. There was no other card of any type that was at Keilor Plains Station on 16 Mar 2016 and Greensborough Station on 3 May 2016 and touched on before 7am at Rosanna Station on 4 May 2018. So these three tweets would have been sufficient to identify Mr Carbines even if he was travelling on an ordinary card.  Indeed, only one other card matched the first two events, using only the dates without detailed times.

This is consistent with other results on location data showing that about 3 points of information are typically sufficient for re-identification, and that precise information is generally not necessary \cite{de2013unique}.

\section{Uniqueness in Transport Data}\label{sec:uniqueness}
Whilst it is clear that this particular data should never have been released in the form it was released in, we can analyse how unique such information is and evaluate whether such data could ever be released in raw unit record form. It is rare to have access to such a large, and possibly whole-of-population dataset, to analyse for uniqueness. The implications of how unique such data is are applicable in a much broader context, for example, in what information card holders are receiving in smart ticketing schemes that permit payments using credit and debit cards.

It is important to note that financial, reputational or physical damage may be suffered by an individual who is included in a unit level record release, without uniqueness. For instance, if an individual is known to match one of a small set of records (but not a unique record), and all of those records involve a sensitive travel event, then an attacker can infer that the sensitive event applies to that individual. Furthermore, large detailed datasets can be used to link between datasets with sensitive attributes and PII, thereby unlocking re-identification, even when absolute uniqueness is absent. %

\subsection{Methodology}
Our approach was similar to that of the unicity measure described by de~Montjoye \textit{et al.}~\cite{de2013unique}. Originally, the data was released in separate files for touch-on and touch-off events,  then  further divided into samples, between 0 and 9, and into separate weeks and years. We reorganised this to put the events for each card into the same record.  We constructed an object  that represented each card, within which was a list of events. Those events could be touch-on or touch-off events, or both, depending on which files were selected to be loaded into the program.

It would be infeasible to evaluate every possible combination of events for every individual for uniqueness. As such, we evaluated the average level of uniqueness by randomly sampling events from within the set of events associated with each card and then determining whether the sampled set of events was unique across all events in the dataset. As such, if we were examining sets of events with a cardinality of 3 we would sample three events at random from each card's set of events. For each set we would then compare those 3 events across all other events for every card in the dataset to determine if they were a unique set of events and therefore the card holder's travel pattern was unique. If a card did not have sufficient events to fill the set cardinality, we selected as many events as possible, aware that this would likely lead to a lower level of uniqueness.

Additionally, we evaluated whether the inclusion of the location information about the touch on or off location had a significant impact on uniqueness---with the expectation that it would. To determine the effect of releasing the dataset after truncation of the time, we evaluated each event with five different granularities of time, as shown in Table~\ref{tab:time_granularity}.

\begin{table}[ht]
\centering
\begin{tabular}{|c|c|}
 \hline
 \emph{Time Granularity} & \emph{Description}  \\
 \hline
 exact & To the second timings as found in the dataset  \\
\hline
 zeroSeconds & Seconds are truncated to zero, leaving time to the minute  \\
\hline
 nearestFiveMinutes & Times are rounded up or down to their nearest five minute interval  \\
 \hline
 zeroMinutes & Minutes and seconds are truncated to zero, leaving time to the hour \\
 \hline
 zeroHour & Hours, minutes, and seconds are truncated to zero, leaving time to the day  \\
\hline
\end{tabular}
\caption{Description of time granularity.}
\label{tab:time_granularity}
\end{table}

\subsection{Determining Uniqueness}
For each event an event signature is created that consists of the time at the respective granularity and if appropriate the location string. This signature acts as a key for the map, within which we store all cardIds that have an event matching that key. The bin of the map is an object itself, a reference to which is held by any card that has a corresponding event matching the event signature.

Each card's objects are instantiated with the events for that card, permuted.
  The full calendar is then populated with all events, whilst simultaneously keeping a reference to the first $n$ events for a particular card, where $n$ is the set cardinality we are analysing. For each card we extract a list of event bins up to the set cardinality. These event bins provide a global view of the calendar for matching cards with those event signatures. We run a simple uniqueness check to determine if there are any other cards that appear in all the selected bins. If not, those $n$ events are  considered unique.

\subsection{Analysis}
The results shown in this section are for touch-on events only, though we  conducted the same analysis for just touch-off events, and for both touch-on and off events. We chose to focus on the touch-on only events because they provided the best coverage across multiple forms of transport. For example, most tram trips do not have an explicit touch-off event associated with them, since it is not required. However, all metro train trips should have an explicit touch off event. Similar results were seen across the different types of events, with the general trend being that the more data that is available the greater the level of uniqueness. This last point is a result of having more cards that fill the set cardinality, for example, some travellers who travel only on trams will only touch on once when activating a monthly or yearly pass. Whilst not strictly in compliance with the rules of the ticketing system, it is generally not something that results in a fine. As such, when looking at a week or a month, that traveller may only have 1 event. Whereas, when looking at a complete year, they could have 12 events, and therefore have more data points to be evaluated on. However, it should be noted, that such a traveller will in general be harder to find due to the unpredictability of their touch-on events and therefore they will present a greater challenge to a targeted attack due to the sparsity and inconsistency of their touch-on events.

\begin{figure}[!ht]
     \centering
     \begin{subfigure}[b]{0.48\textwidth}
         \centering
         \includegraphics[width=\textwidth]{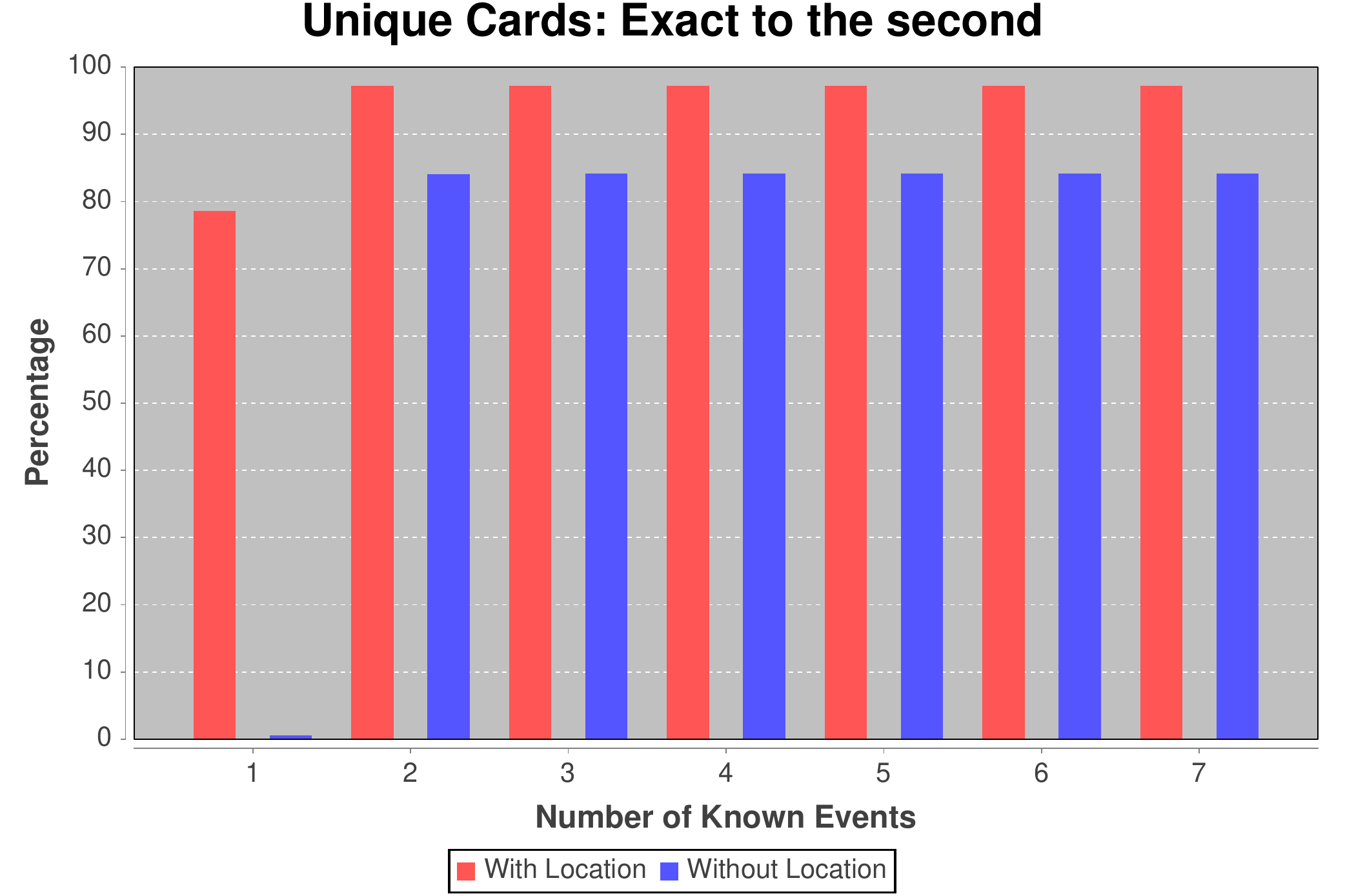}
         \caption{Unique Cards - exact time to the second}
         \label{2017_week40_touchon_exact}
     \end{subfigure}
     \hfill
    \begin{subfigure}[b]{0.48\textwidth}
         \centering
         \includegraphics[width=\textwidth]{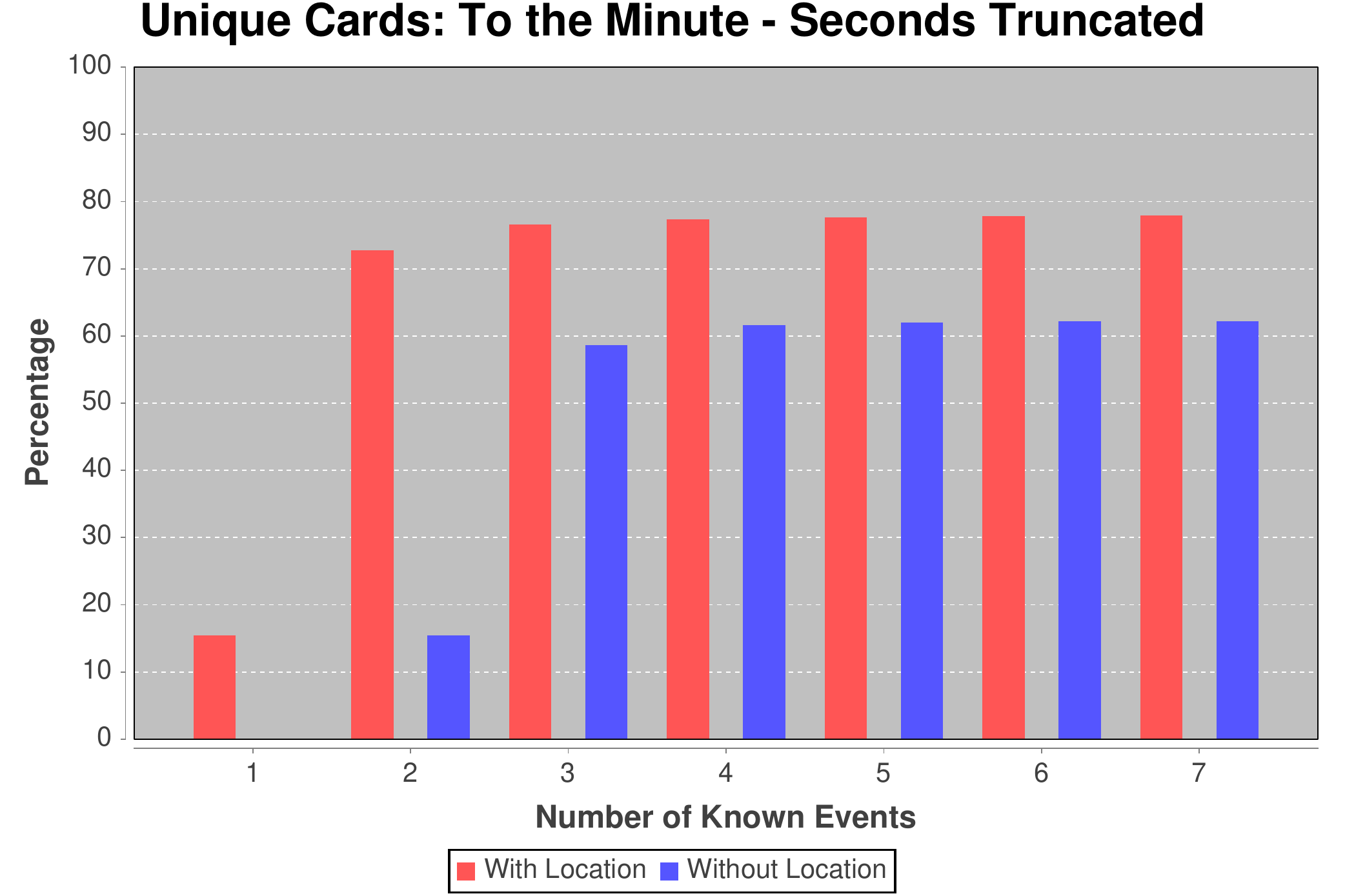}
         \caption{Unique Cards - truncated to minutes}
         \label{2017_week40_touchon_zeroseconds}
     \end{subfigure}

     \begin{subfigure}[b]{0.48\textwidth}
         \centering
         \includegraphics[width=\textwidth]{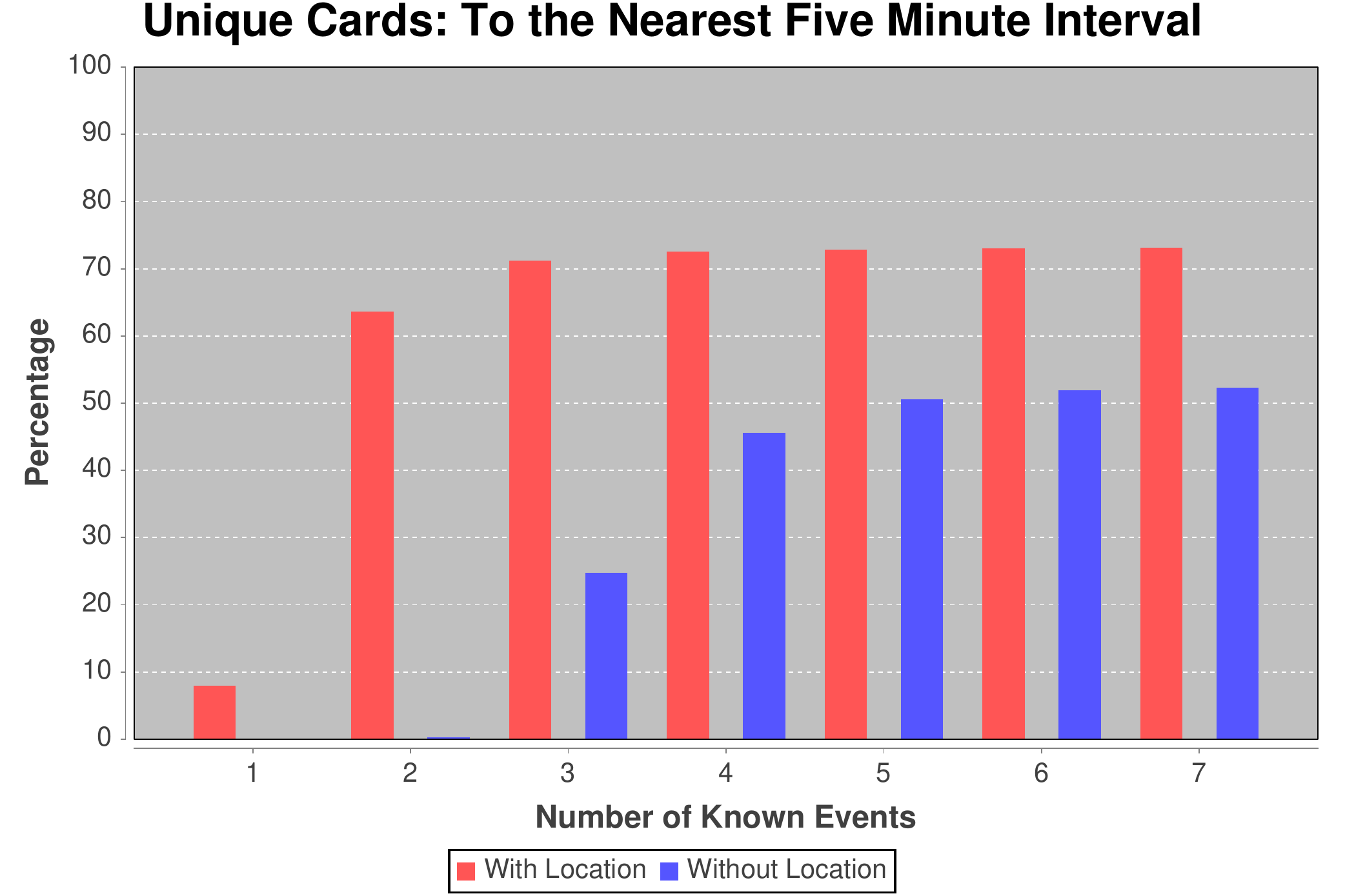}
         \caption{Unique Cards - truncated to nearest 5 minutes}
         \label{2017_week40_touchon_near5}
     \end{subfigure}
     \hfill
    \begin{subfigure}[b]{0.48\textwidth}
         \centering
         \includegraphics[width=\textwidth]{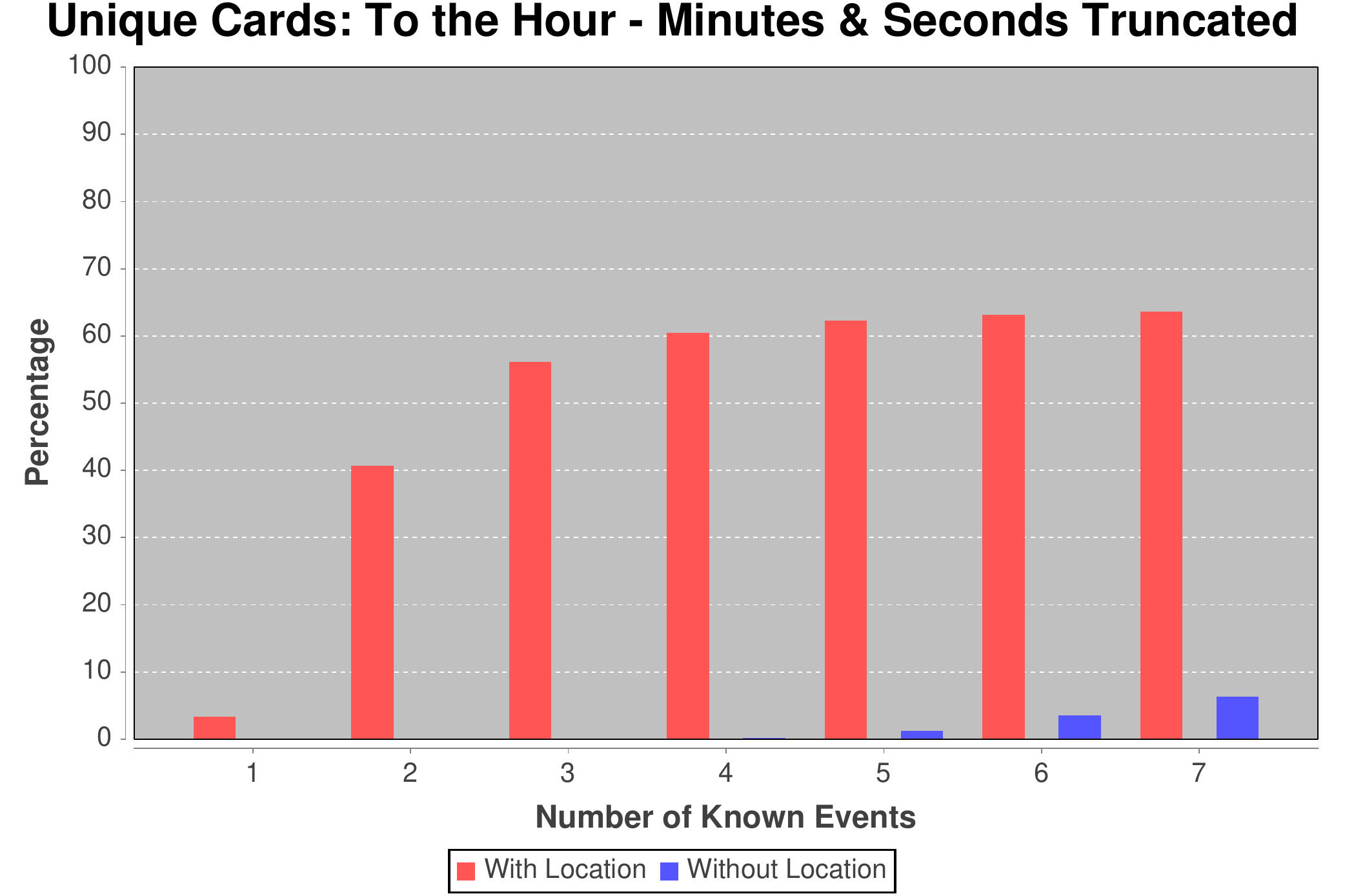}
         \caption{Unique Cards - truncated to hours}
         \label{2017_week40_touchon_zerominutes}
     \end{subfigure}

     \begin{subfigure}[b]{0.48\textwidth}
         \centering
         \includegraphics[width=\textwidth]{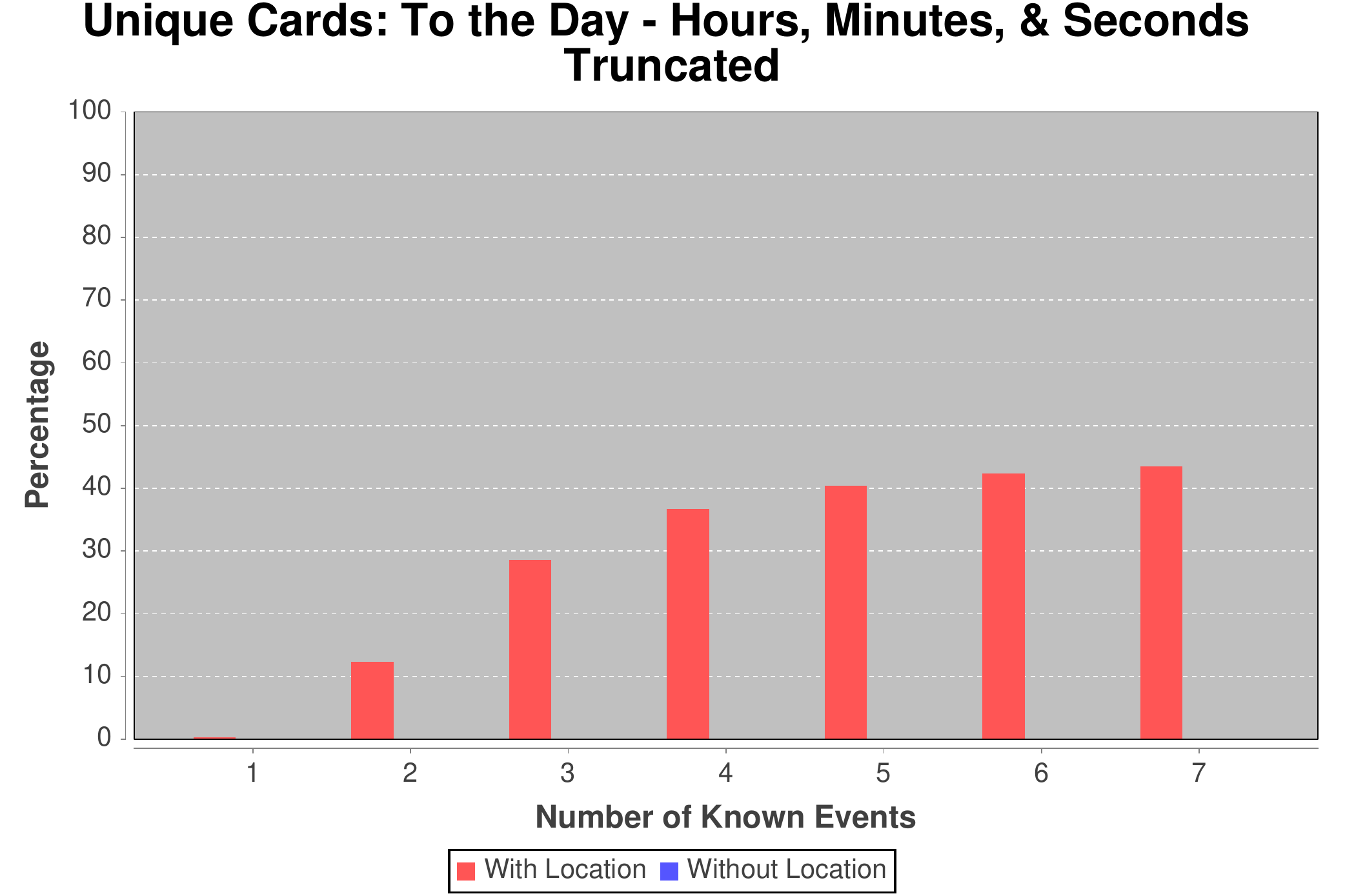}
         \caption{Unique Cards - truncated to day}
         \label{2017_week40_touchon_zerohours}
     \end{subfigure}

    \caption{Analysis of unique cards from a single week of data.}
    \label{fig:week_events}
\end{figure}

Figure~\ref{fig:week_events} shows the analysis results for  a single week (week 40) of touch-on events in 2017. This week was picked at random from the full set of weeks available in 2017. The graphs show the percentage of cards that had unique sets of events out of the cards that had at least one corresponding event in the analysed period. Figure~\ref{2017_week40_touchon_exact} shows the percentage of cards that were unique based on the exact touch on times, both with and without location. It is clear to see that the exact time provides significant uniqueness after only two randomly selected events. With location the uniqueness is close to 80\% with just a single event. This stands to reason, given that the chances of two people touching on at the exact same second on the same tram or at the same station is low, not least because of the finite number of Myki ticket devices on which someone could touch on. The uniqueness rapidly plateaus as the number of events increases. This will in part be down to the number of cards that only have a single touch-on event or only a handful of touch-on events. These could be cards from season ticket holders or tourists who have low frequency usage patterns. The plateau effect is observed in the other graphs and wider analysis, with the difference being after how many events the plateau begins.

As can be seen from Figure~\ref{2017_week40_touchon_zeroseconds}, whereby the seconds have been truncated to zero, there is still a high rate of uniqueness after two events when including location, and over 50\% uniqueness from three events without location. This is an important finding because the granularity of touch events shown by Myki machines is at the level of minutes not seconds. It is only on the website that results to the second are provided. Discovering that a high rate of uniqueness is discoverable even at the minute level indicates that even a person who does not register their card, and therefore does not have access to per second records, is still likely to be able to find themselves within the dataset from only 3 events.

Even binning to the nearest five minute interval, as shown in Figure~\ref{2017_week40_touchon_near5}, shows a greater than 60\% level of uniqueness with two events when location is known. Without location it is above 50\% after five events. Even when truncating the minutes and seconds, leaving just hour---see Figure~\ref{2017_week40_touchon_zerominutes}---the power of location to identify cards is clear, with above 50\% being unique with three events, and above 60\% uniqueness with four events. We would not expect time alone to be particularly unique, purely because there is so little entropy in the hour travelled. It was therefore surprising to see some cards starting to demonstrate uniqueness even without location after five events. Whilst a small number, it indicates that there are unusual travelling patterns, for example, very early or very late, that can still be unique even when removing minutes and seconds from the event time.

Figure~\ref{2017_week40_touchon_zerohours} shows the results from truncating hours as well as minutes and second, effectively providing just day as the time parameter. Not surprisingly, time alone does not display any uniqueness, however, location exhibits an almost 30\% level of uniqueness after just three events.

\begin{figure}[!ht]
     \centering
     \begin{subfigure}[b]{0.48\textwidth}
         \centering
         \includegraphics[width=\textwidth]{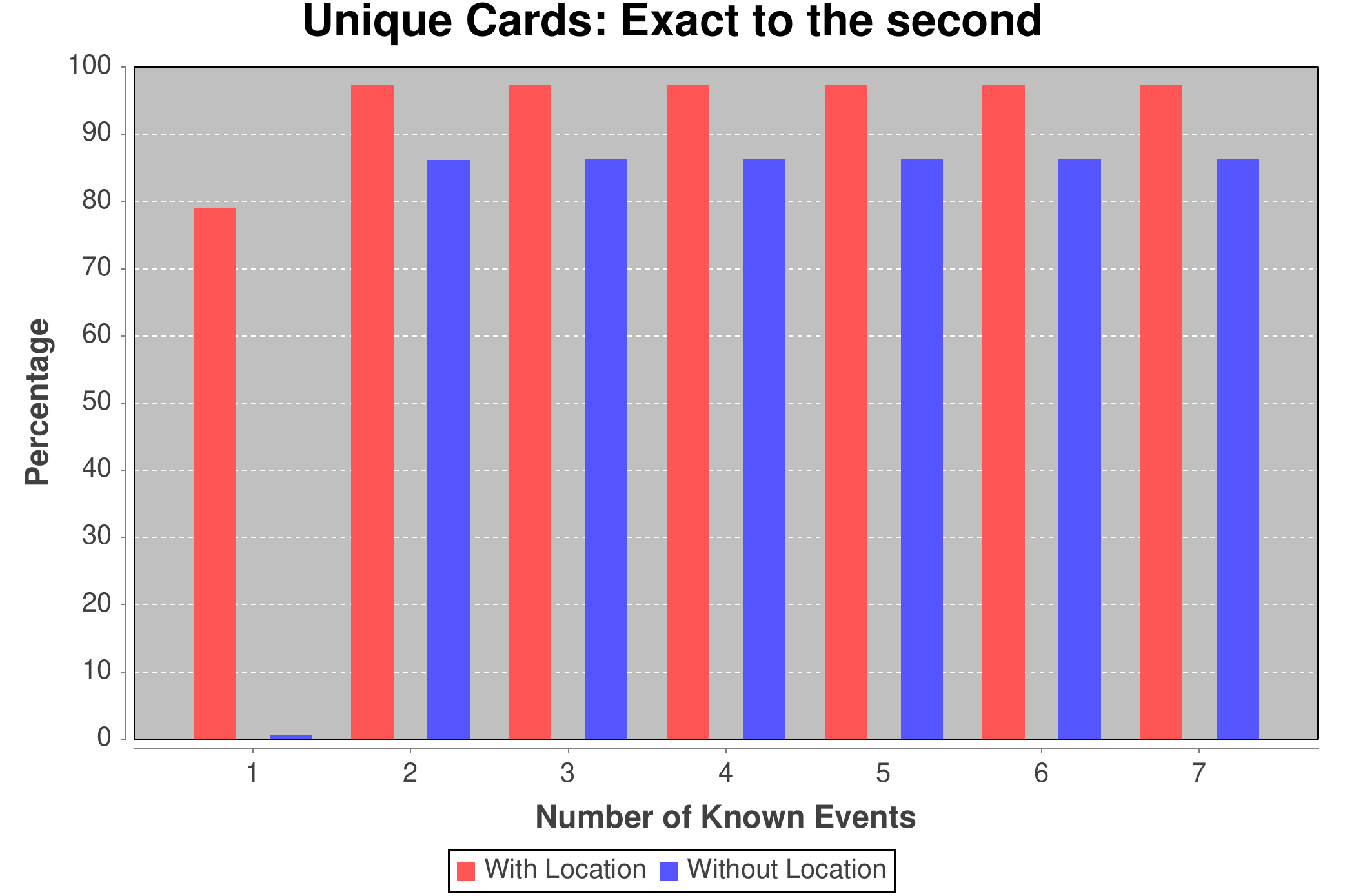}
         \caption{Unique Cards - exact time to the second}
         \label{2017_fullmonth_touchon_exact}
     \end{subfigure}
     \hfill
    \begin{subfigure}[b]{0.48\textwidth}
         \centering
         \includegraphics[width=\textwidth]{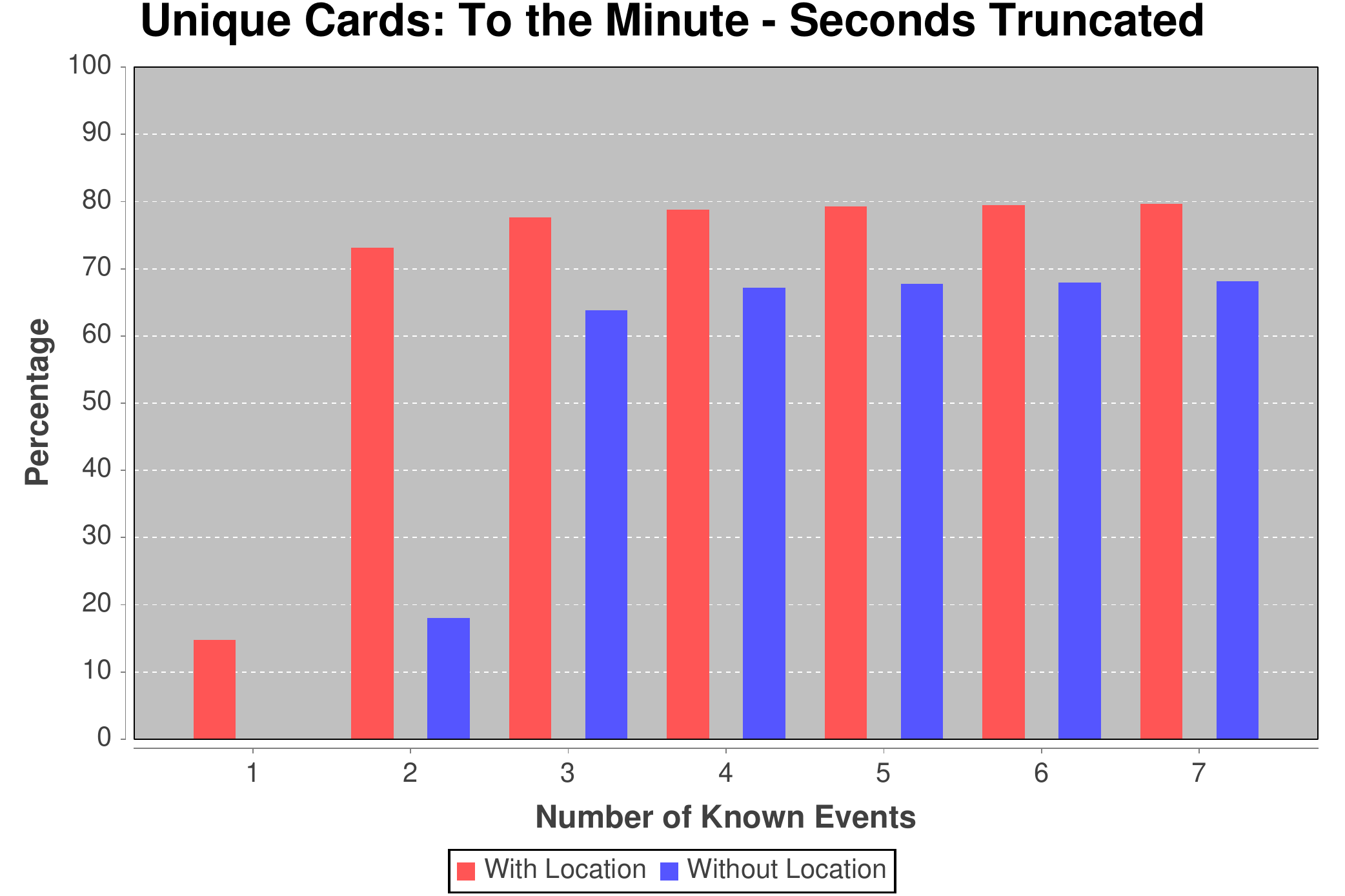}
         \caption{Unique Cards - truncated to minutes}
         \label{2017_fullmonth_touchon_zeroseconds}
     \end{subfigure}

     \begin{subfigure}[b]{0.48\textwidth}
         \centering
         \includegraphics[width=\textwidth]{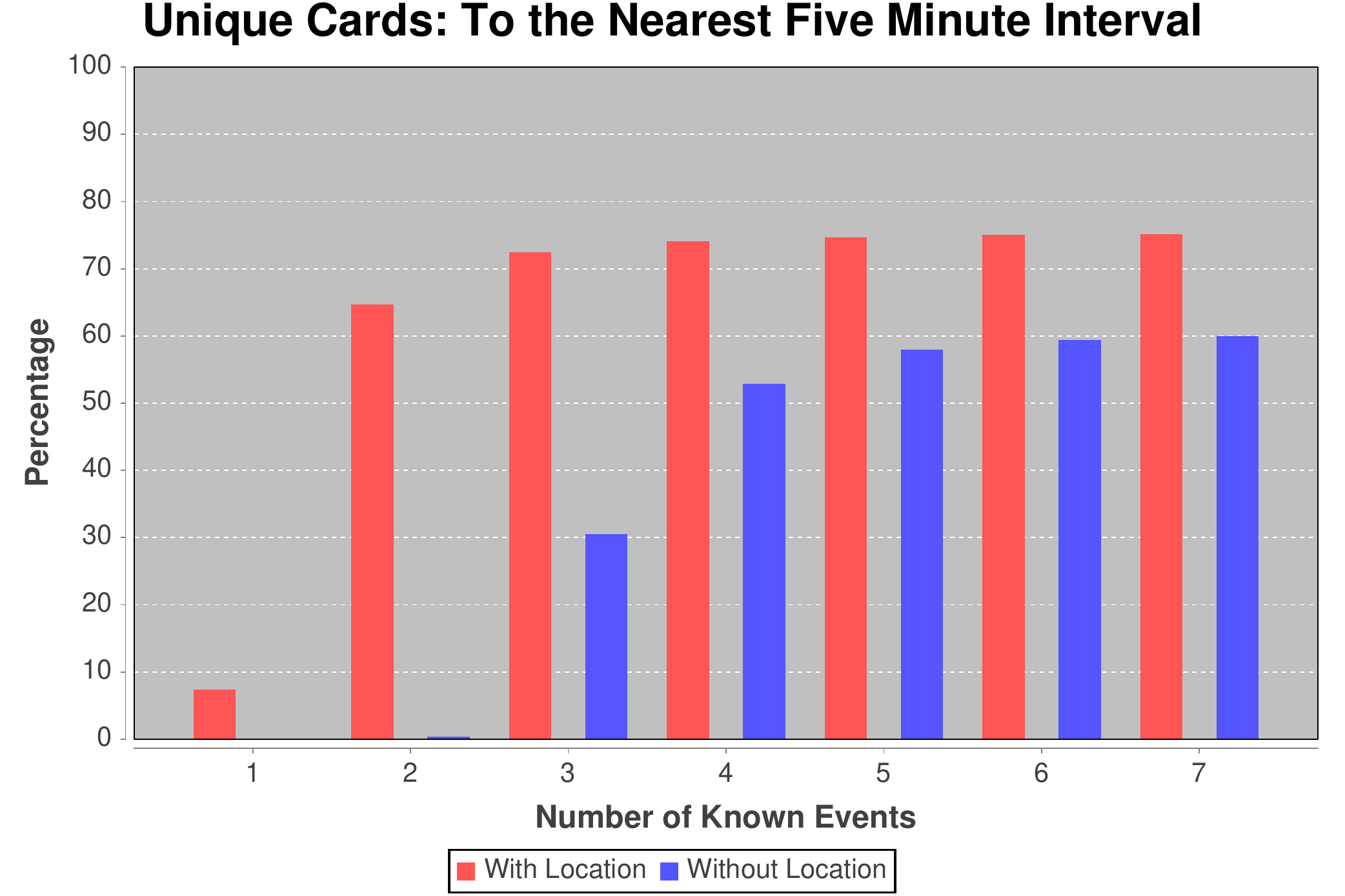}
         \caption{Unique Cards - truncated to nearest 5 minutes}
         \label{2017_fullmonth_touchon_near5}
     \end{subfigure}
     \hfill
    \begin{subfigure}[b]{0.48\textwidth}
         \centering
         \includegraphics[width=\textwidth]{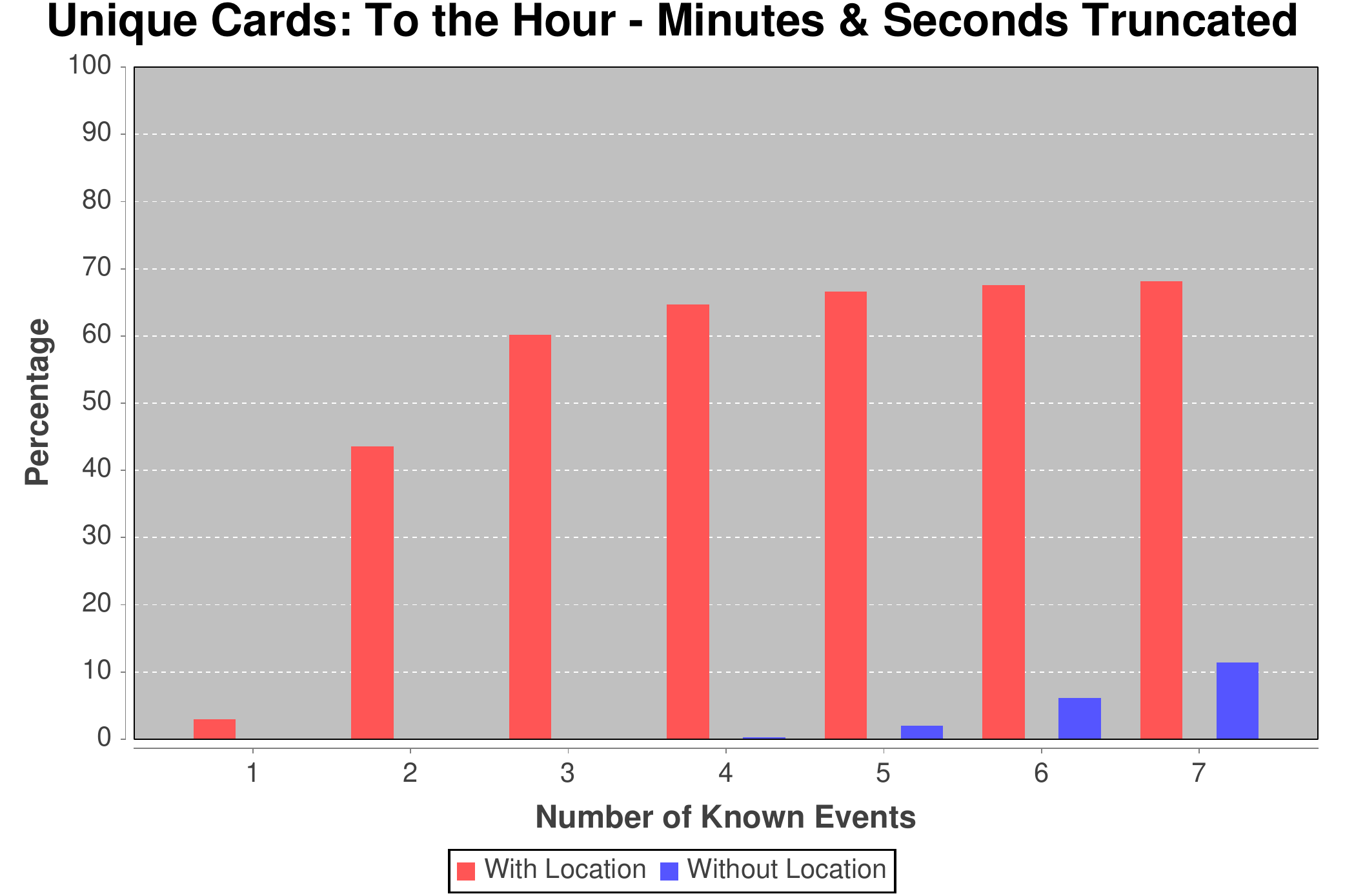}
         \caption{Unique Cards - truncated to hours}
         \label{2017_fullmonth_touchon_zerominutes}
     \end{subfigure}

     \begin{subfigure}[b]{0.48\textwidth}
         \centering
         \includegraphics[width=\textwidth]{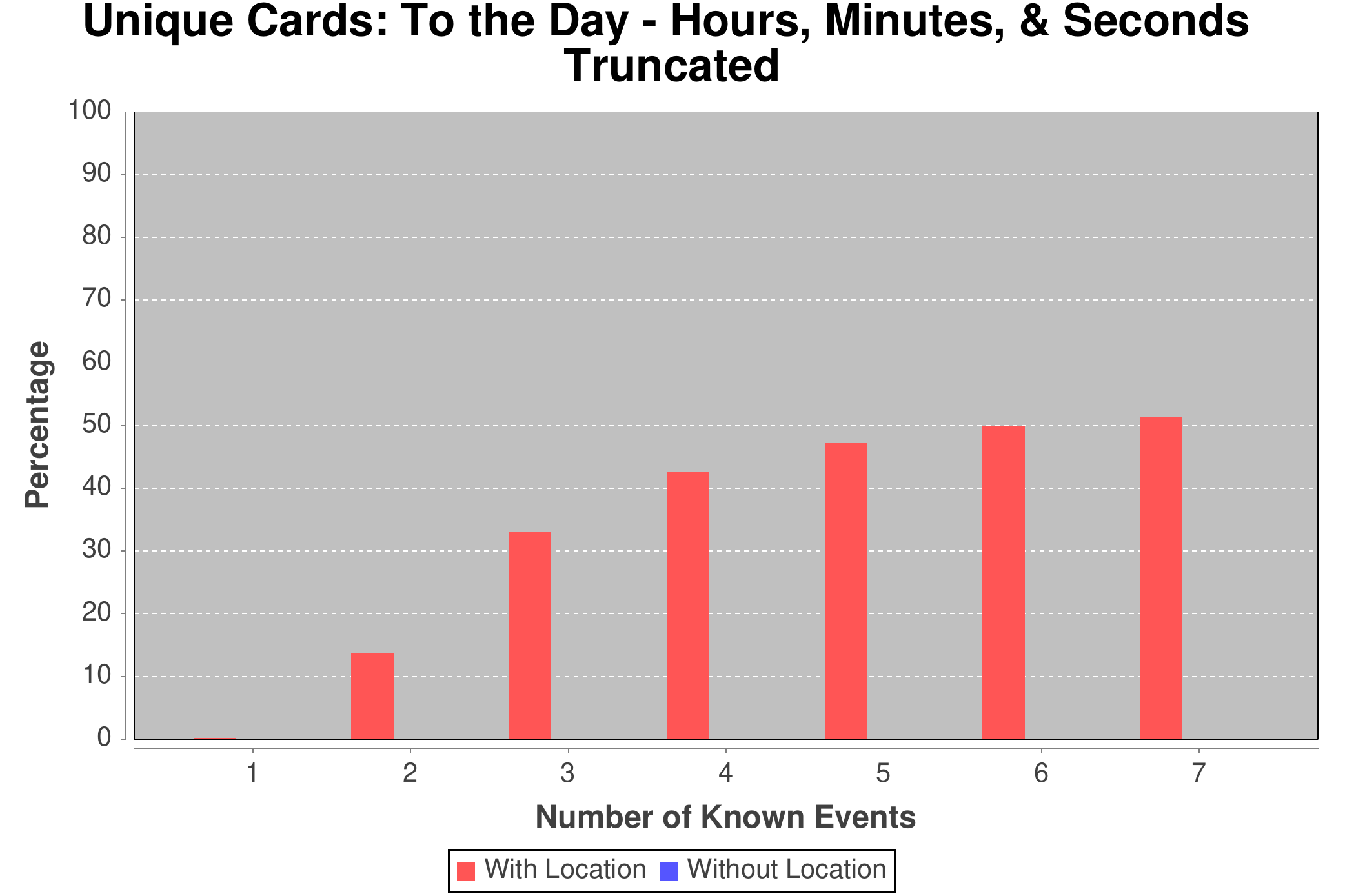}
         \caption{Unique Cards - truncated to day}
         \label{2017_fullmonth_touchon_zerohours}
     \end{subfigure}

    \caption{Analysis of unique cards from a month of data.}
    \label{fig:month_events}
\end{figure}

\begin{figure}[!ht]
     \centering
     \begin{subfigure}[b]{0.48\textwidth}
         \centering
         \includegraphics[width=\textwidth]{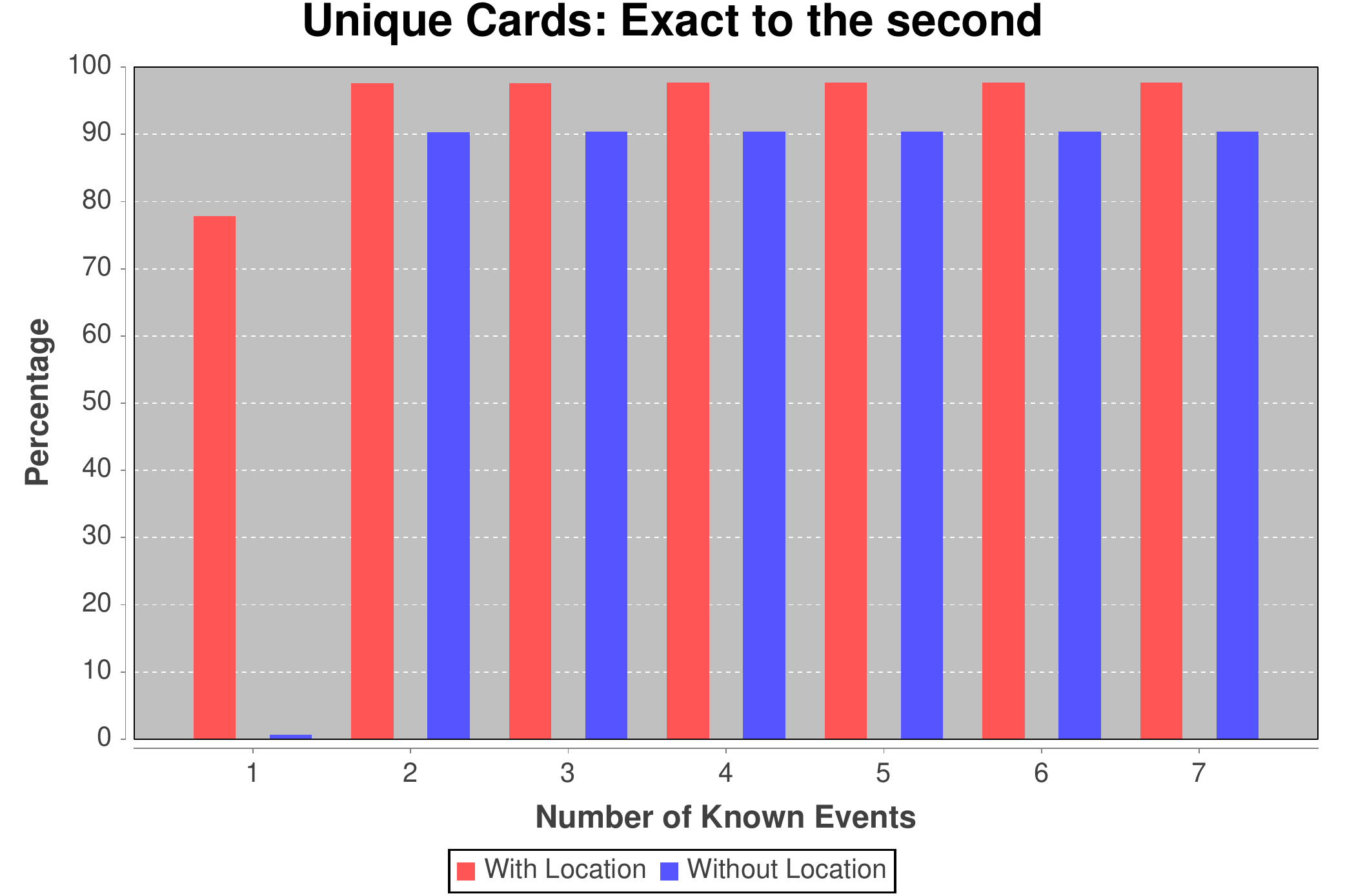}
         \caption{Unique Cards - exact time to the second}
         \label{2017_fullyear_touchon_exact}
     \end{subfigure}
     \hfill
    \begin{subfigure}[b]{0.48\textwidth}
         \centering
         \includegraphics[width=\textwidth]{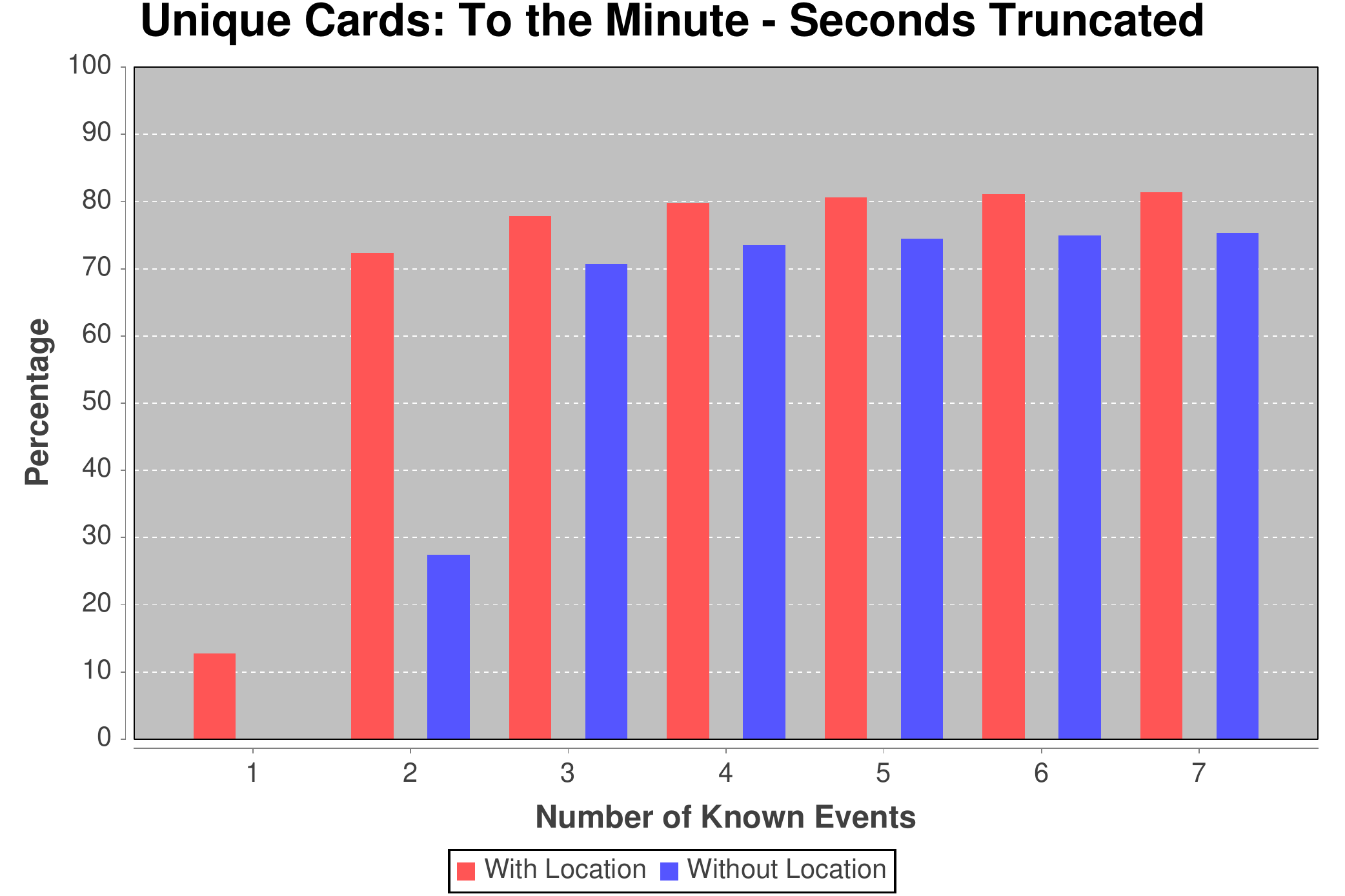}
         \caption{Unique Cards - truncated to minutes}
         \label{2017_fullyear_touchon_zeroseconds}
     \end{subfigure}

     \begin{subfigure}[b]{0.48\textwidth}
         \centering
         \includegraphics[width=\textwidth]{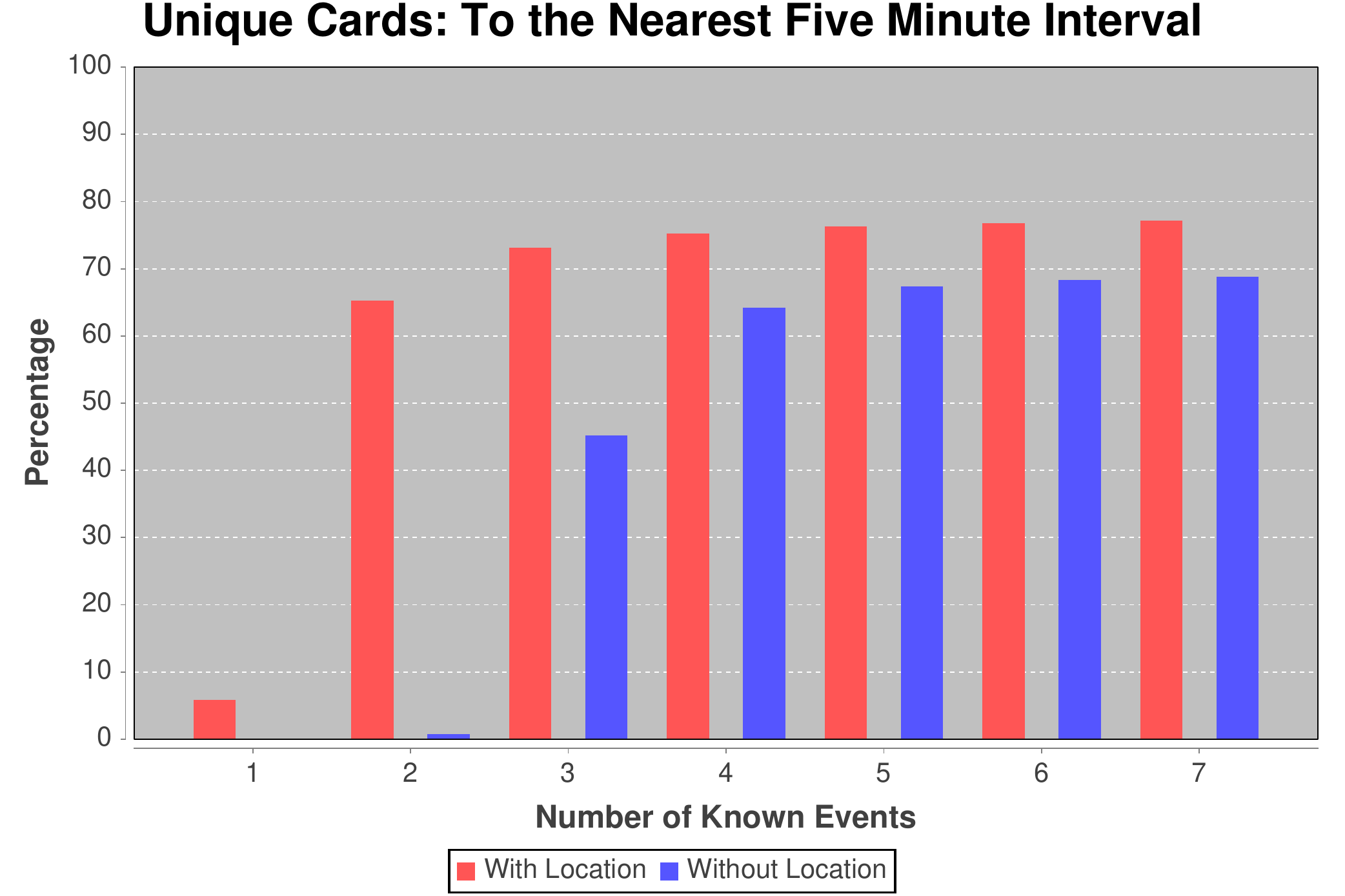}
         \caption{Unique Cards - truncated to nearest 5 minutes}
         \label{2017_fullyear_touchon_near5}
     \end{subfigure}
     \hfill
    \begin{subfigure}[b]{0.48\textwidth}
         \centering
         \includegraphics[width=\textwidth]{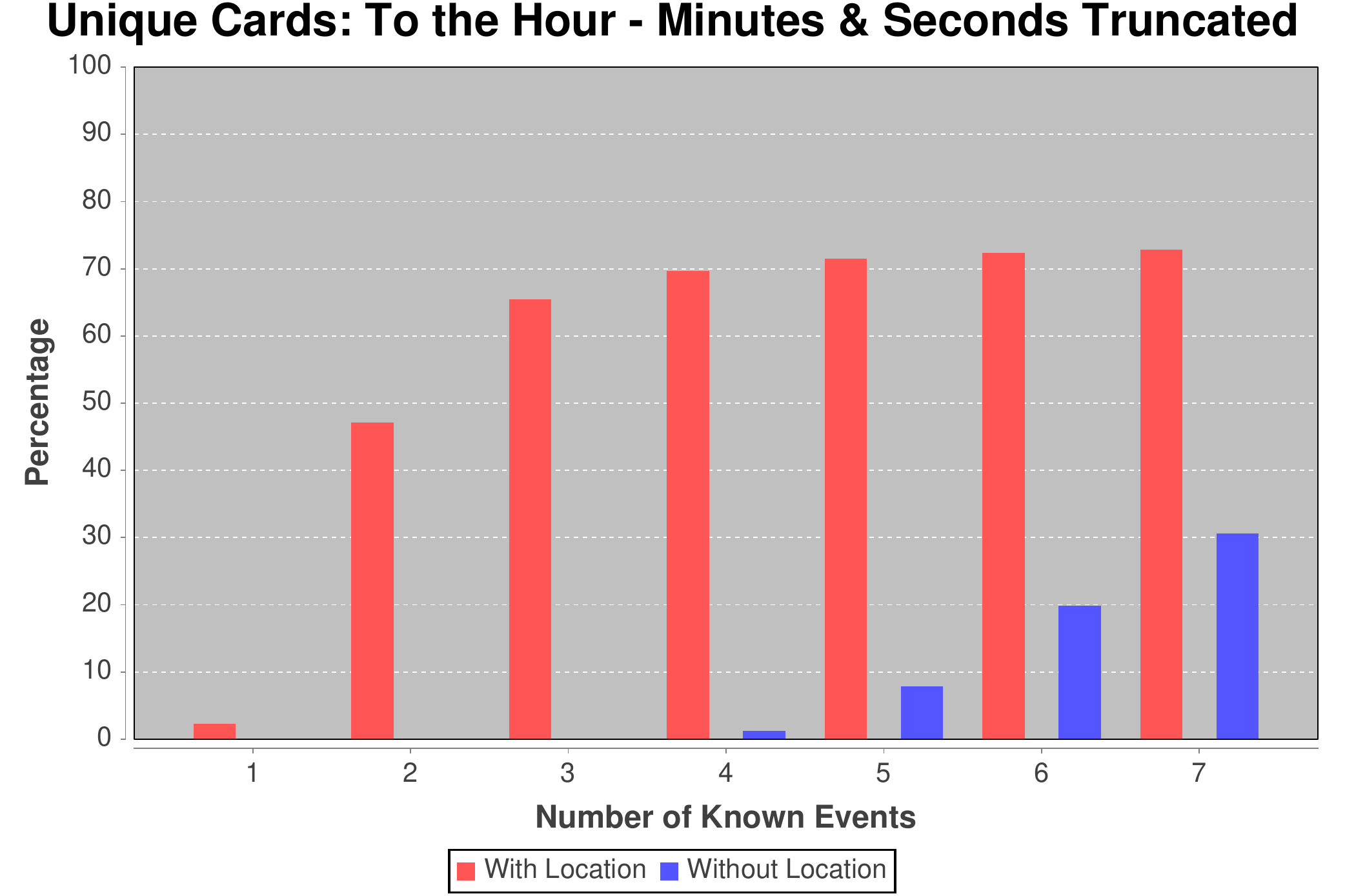}
         \caption{Unique Cards - truncated to hours}
         \label{2017_fullyear_touchon_zerominutes}
     \end{subfigure}

     \begin{subfigure}[b]{0.48\textwidth}
         \centering
         \includegraphics[width=\textwidth]{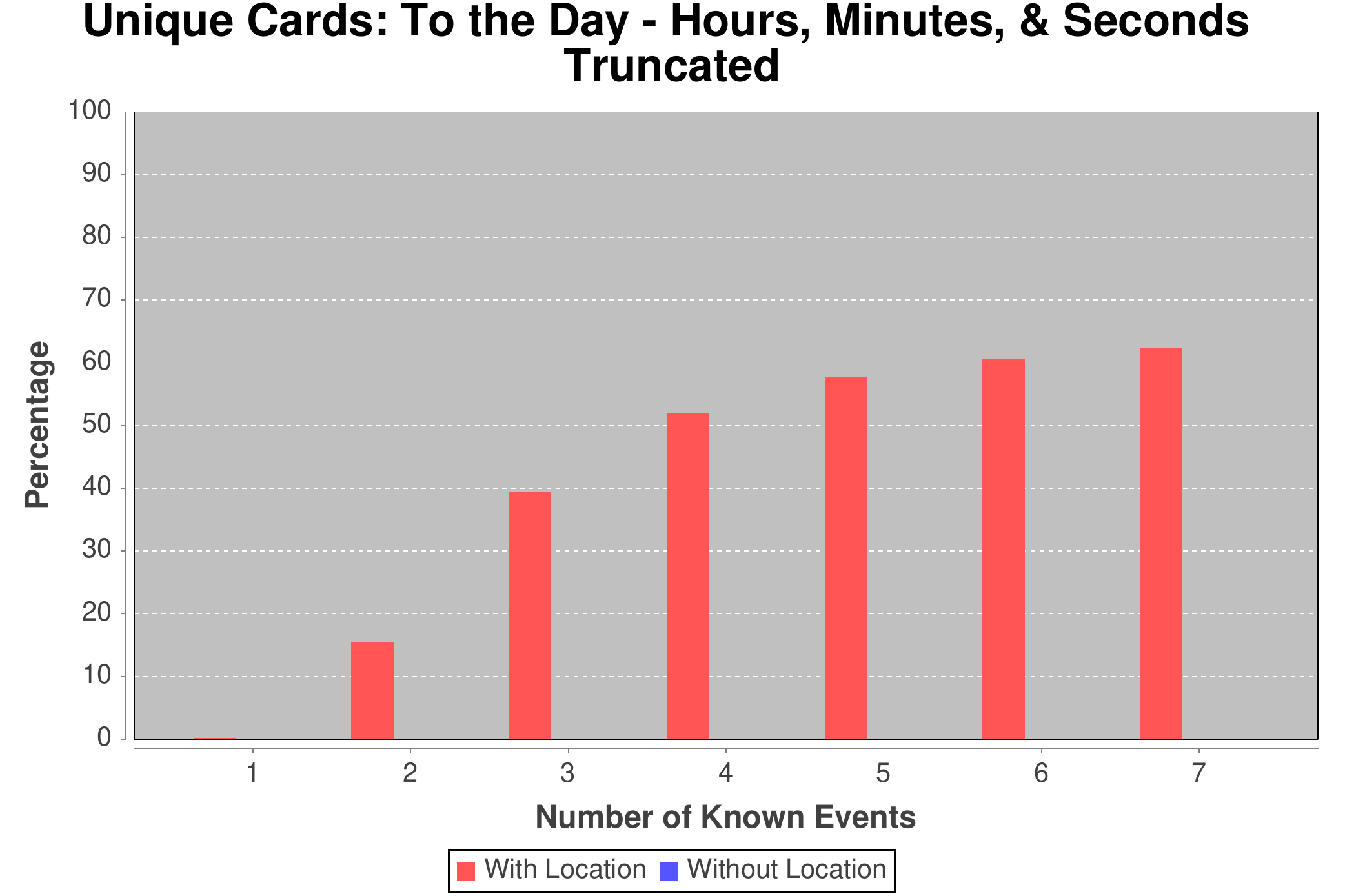}
         \caption{Unique Cards - truncated to day}
         \label{2017_fullyear_touchon_zerohours}
     \end{subfigure}

    \caption{Analysis of unique cards from a year of data.}
    \label{fig:year_events}
\end{figure}

Figures~\ref{fig:month_events} and \ref{fig:year_events} show the same breakdown of time granularities but for a period of one month and one year respectively. We see the same trends as described above, but with a greater level of uniqueness the more data that is available. For example, with one year of data to select from, over 90\% of cards are unique based on exact time of two random events. For one month it is over 85\% for the same set size, and a little over 80\% for a single week.

\clearpage

\section{How Could this Open Data Release Have Been Done Better?}\label{sec:better}
In 2016, Transport for NSW released as open data the tap-on and tap-off events for their public transport ticketing system, Opal.  Superficially, this release seems similar to the Myki release---the ticketing systems record similar information and the raw datasets must have been very similar.  However, the NSW release was much more careful to protect privacy.

First, the \emph{technical} mechanisms were far more privacy preserving.  Rather than release longitudinal data for each card, the Opal data mentions only the total number of touch-on or touch-off events at each place and time.  Even if you know several precise events for a person, there is no way of retrieving other events on the same card.  These totals are aggregated into quarter-hour time blocks, and then mechanisms from differential privacy~\cite{dwork2006calibrating,dwork2014algorithmic} are applied to obscure the exact totals. Crucially, differential privacy provides a specific security property such that data releases that preserve differential privacy effectively protect against attackers with a well-defined range of knowledge and capabilities.

Second, the \emph{procedures} were much more careful.  Transport for NSW engaged CSIRO/Data61 to process the dataset, and contracted us to examine the security of what they had done before it was published online.  Our conclusion was that the privacy protection was sufficient, though we made suggestions for future releases including stronger differential privacy protection.  Data61's detailed explanation of their mechanisms\footnote{\url{https://arxiv.org/abs/1705.05957}}, our analysis\footnote{\url{https://arxiv.org/abs/1704.08547}} and Data61's response\footnote{\url{https://arxiv.org/abs/1705.08994}} are all openly available online.  This means that details about the privacy protection mechanisms are open, too, allowing for further analysis and discussion that can inform future data sharing.

It is important to note that these techniques for privacy preservation do remove detail and connections---there is no free lunch.  In the Myki data, it is possible to investigate trips and habits of people across different days and locations.  In the Opal data, trip or journey analysis is significantly diminished.  This reduction of information is generally inevitable when releasing a detailed dataset, however useful data can be shared while protecting privacy via careful technical and procedural measures.

We recommend:
\begin{itemize}
    \item
not releasing detailed (\textit{i.e.,} longitudinal) unit-record level data; \item applying differential privacy techniques to aggregate data;
\item openness about the privacy-protection mechanisms; and \item examination by independent experts before release.
\end{itemize}

\section{Opinion: What Needs to Change in Australian Law and Policy?}\label{sec:discussion}
\subsection{The Connection with Federal Open Data Breaches}
This is the second time our team has re-identified individuals in an open  dataset published by government in
the mistaken belief that it had been securely de-identified.
In 2016 the Federal Department of Health released  10\% of the Australian population's complete Medicare Benefits Scheme and Pharmaceutical Benefits Scheme billing data for the 30 year period 1984--2014. If an Australian was unlucky enough to be in the sample, their entire medical billing records for that period were made available online, publicly, without even registration required for download.

The then Health Minister Sussan Ley admitted that doctors could be identified but reportedly claimed that ``No patient information has been compromised,''\footnote{\url{https://www.abc.net.au/news/2016-09-29/medicare-pbs-dataset-pulled-over-encryption-concerns/7888686}} though complete data for nearly 3 million patients was posted on the Internet.  The Office of the Australian Information Commissioner officially reported that ``the patients in the dataset are not reasonably identifiable for the purposes of the Privacy Act,''\footnote{\url{https://www.oaic.gov.au/privacy/privacy-decisions/investigation-reports/mbspbs-data-publication/}} though we had demonstrated confident re-identification of several patients~\cite{culnane2017health}.  And despite insisting that patients were not identifiable, then Attorney General George Brandis attempted (but fortunately failed) to criminalise further re-identification.\footnote{\url{https://www.aph.gov.au/Parliamentary_Business/Bills_Legislation/Bills_Search_Results/Result?bId=s1047}}
As far as we know, none of the affected people have been notified.

The Australian government's response to such a serious error in judgement was not to retreat from increased sharing. Instead even greater sharing is being  facilitated, without clear privacy protections, by committing to ``make more public data openly available and support[s] its use to launch commercial and non-profit ventures, conduct research, make data-driven decisions and solve complex problems,''\footnote{\url{http://ogpau.pmc.gov.au/commitment/21-release-high-value-datasets-and-enable-data-driven-innovation}} as part of its Open Government Partnership.

The Victorian government's re-identifiable data release is a predictable consequence of the Australian government's 
refusal to deal with the MBS-PBS identifiable data release. 
The mistakes will keep being repeated until the failings are acknowledged and addressed in a transparent manner.

\subsection{Open Government vs. Open Data} \label{sec:OpenGovtvsOpenData}

Open data  releases are framed as necessarily beneficial to society and are often made under the guise of open government.
Genuine openness \emph{about government} provides clear benefit with minimal potential privacy cost.
President Obama's Memorandum on \emph{Transparency and Open Government}~\cite{obama2009memo} was clear in its goal of increasing transparency of government, stating ``Executive departments and agencies should harness new technologies to put information about their operations and decisions online and readily available to the public." That worthy goal has been distorted to instead provide greater transparency on the population itself, often at a cost to citizens' human rights.

Government must strike a balance between cost to individual human rights of sharing identifiable data, legitimate and consented use of citizen data for true societal benefit, and
data shared due to lobbying by  
commercial or professional self-interest.  
Strong cases must be made for collecting, retaining and sharing sensitive data. Where the case is clearly in the public good, explicit consent and strong technical privacy protections must be employed and mandated by law. Over a decade of seminal papers on the fallacy of de-identification and global news of failed data sharing experiments~\cite{sweeney2000uniqueness,narayanan2008robust,narayanan2009anonymizing,de2013unique,Pandurangan2014taxi,culnane2017health,ovic2018} prove that the \emph{status quo} is untenable.

\subsection{Non-open Sharing of De-identified Data}
The publication of inadequately de-identified open data is
probably only the most visible part of a much larger problem of non-open sharing of data, without the consent of the data
subjects, on the basis that it has been de-identified and therefore lies outside the constraints of the Commonwealth \emph{Privacy Act 1988} or state equivalents.  When that data is made openly available we can demonstrate that
it is, in fact, identifiable.  When it circulates without
scrutiny, it is impossible to tell.

Some examples of data sharing are examined in the media or advertised publicly.
The National Australia Bank shared ``de-identified customer transaction data from all NAB account holders,'' representing that the data was ``not considered to be personal information subject to Australia’s privacy laws.''\footnote{\url{https://www.abc.net.au/news/2019-03-05/sportsbet-documents-reveal-millions-spent-on-marketing/10833196}}  Telstra has sold its customers' location information through their Location Insights program, claiming ``The insights are disclosed in an aggregated and anonymised form and do not include any personally identifiable customer data.''\footnote{\url{https://locationinsights.telstra.com/}}  Australian company Nostradata reportedly partners with global data broker IMS Health (now IQVIA) to share ``NostraData’s de-identified pharmacy dispensary information.''\footnote{\url{https://www.pharmacydaily.com.au/news/ims-nostradata-deal/48458}}
My Health Record patients are told ``You may also choose to share de-identified data from your My Health Record for research purposes. When health data is de-identified, it means you are anonymous and it can’t be traced back to you.''\footnote{\url{https://www.myhealthrecord.gov.au/for-you-your-family/howtos/choose-how-your-data-is-used-for-research}}

In all of these cases (and the others we don't know about), there is room for scepticism about the success of the de-identification, particularly when no specific security property is being guaranteed.  We do not know of any genuine, open, expert scrutiny of any of this data.  If it were in fact easily re-identifiable, we do not see how this fact would come to light.

The recently-passed \emph{Consumer Data Right} adds to the ambiguity by treating de-identification as equivalent to deletion, and allowing the data recipient to decide which to use. Furthermore, it creates a modified definition of de-identification from the one found in the \emph{Privacy Act 1988}, using the word \emph{relates} instead of \emph{about}, claiming in the Explanatory Memorandum that it broadens the concept of identifiability, although it is not clear from the dictionary definitions of those words how that broadening is achieved.

For 10\% of Australians, many of the above datasets are very easily linked with their already-published MBS-PBS data.  This is particularly relevant for health data and financial transactions data, and even more so if it goes back to 2014.

The ACCC's report on digital platforms  recommends strengthening the Commonwealth \emph{Privacy Act 1988}.\footnote{\url{https://www.accc.gov.au/focus-areas/inquiries/digital-platforms-inquiry}}
\begin{quotation}
the Government could consider whether the Privacy Act should set out additional protections or requirements for the de-identification, anonymisation, or aggregation of personal information. These are all ways that entities may remove personally identifying information so that the information is no longer within the scope of the Privacy Act. However, there are increasing risks that such information may become re-identified as more information becomes available, multiple datasets are combined, and advances in data analytics are made.
\end{quotation}

We are not lawyers, so we will not comment on whether the law needs to be strengthened or simply more consistently complied with, at a state or federal level.  Certainly the practice of publishing or sharing  identifiable personal information, while claiming it is ``de-identified'' and therefore outside the constraints of the Privacy Act, must stop.

\section{Conclusion} \label{sec:conclusion}
Ordinary travellers are very easily and confidently identifiable from the published Myki data.  It takes about 3 points with dates to identify a complete stranger, but only one with an exact time to identify someone who travelled with you.  This then allows the retrieval of all their travel records for months or years.

Future releases should follow the recommendations in Section~\ref{sec:better}.

A more serious problem is re-identifiable datasets that have been shared without the affected people ever finding out---this makes it impossible for them to protect themselves or to make
fully-informed decisions about further sharing.

\bibliographystyle{unsrt}
\bibliography{full}

\clearpage

\begin{partialdetails}

\appendix
\section{15 Other Events Matching Anthony Carbines's Tweets}

\begin{longtable}{p{2cm}p{1.5cm}p{12cm}}
{\bf Date} & & {\bf Tweet /  event in data} \\
\hline
18 July 2015 &  8:56am  &  ``Sunshine and new seating shelter at Rosanna station. Super train trip to G-Town to see the Catters v Dogs today.'' \\
 & & \begin{minted}[escapeinside=||]{js}
{
  "onDate": ISODate("2015-07-18T08:51:36Z"),
  "onMode": 2,
  "onVid": 0,
  "onParentRoute": "",
  "onRouteId": 20,
  "onStopId": |{\bf 19936}|,
  "offDate": ISODate("2015-07-18T09:33:42Z"),
  "offMode": 2,
  "offVid": 0,
  "offParentRoute": "",
  "offRouteId": 1,
  "offStopId": 64408
}
 \end{minted}
 \\
  & & Perfect match (5 mins).  Note the offStopId differs from regular commute, probably Southern Cross.  \\
  \hline
 \\
2 Sep 2015 & 10:20pm & ``Train graffiti...usually pretty clean trains but not much to look at on Hurstbridge Line tonight...''\\
 & & \begin{minted}[escapeinside=||]{js}
{
  "onDate": ISODate("2015-09-02T21:37:32Z"),
  "onMode": 2,
  "onVid": 0,
  "onParentRoute": "",
  "onRouteId": 1,
  "onStopId": 64403,
  "offDate": ISODate("2015-09-02T22:20:23Z"),
  "offMode": 2,
  "offVid": 0,
  "offParentRoute": "",
  "offRouteId": 20,
  "offStopId": | {\bf 19936 } |
}

\end{minted}
 \\

 & & Perfect match to the minute.  \\
  \hline
 \\\pagebreak
 \hline
8 Oct 2015 & 8:41am & ``8.41 Hurstbridge train from Rosanna cancelled....next at 8.49...'' \\
 & & \begin{minted}[escapeinside=||]{js}
 {
  "onDate": ISODate("2015-10-08T08:36:49Z"),
  "onMode": 2,
  "onVid": 0,
  "onParentRoute": "",
  "onRouteId": 20,
  "onStopId": |{\bf 19936}|,
  "offDate": ISODate("2015-10-08T09:09:41Z"),
  "offMode": 2,
  "offVid": 0,
  "offParentRoute": "",
  "offRouteId": 18,
  "offStopId": 64407
}
\end{minted}
\\
 & &  Perfect match. \\
   \hline
 \\

17 Jan 2016 & 10:24pm & ``Haven't seen this much anticipation since Twiggy Dunne was lining up post the siren. Ah, train other way arrives'' \\
17 Jan 2016 & 10:46pm & ``We are getting off the train as door rooted. Ah, Clifton Hill. What a joke'' \\
17 Jan 2016 & 10:46pm & ``How many doors on a six carriage train?''\\
17 Jan 2016 & 10:51pm & ``Sign of the times. Not really what a train station is all about...'' \\
17 Jan 2016 & 10:59pm & ``Jolimont train dumps us at Cliffy Hill.'' \\
    & & \begin{minted}[escapeinside=||]{js}
{
  "onDate": ISODate("2016-01-17T22:17:03Z"),
  "onMode": 2,
  "onVid": 0,
  "onParentRoute": "",
  "onRouteId": 18,
  "onStopId": 64407,
  "offDate": ISODate("2016-01-17T23:18:31Z"),
  "offMode": 2,
  "offVid": 0,
  "offParentRoute": "",
  "offRouteId": 20,
  "offStopId": |{\bf 19936}|
}
\end{minted}
\\
    & & Consistent. Note late-night arrival. \\
      \hline
 \\\pagebreak
 \hline

10 Mar 2016 & 7:59am &
``Our train driver has a sense of humour;'' \\
     & & \begin{minted}[escapeinside=||]{js}
{
  "onDate": ISODate("2016-03-10T07:47:42Z"),
  "onMode": 2,
  "onVid": 0,
  "onParentRoute": "",
  "onRouteId": 20,
  "onStopId": | {\bf 19936} | ,
  "offDate": ISODate("2016-03-10T08:34:16Z"),
  "offMode": 2,
  "offVid": 0,
  "offParentRoute": "",
  "offRouteId": 1,
  "offStopId": 64403
}
\end{minted}
      \\
     & &  Consistent. \\
           \hline
 \\

30 Apr 2016 & 11:11pm &  ``Catters by 20 goals and last train to Melbs a very satisfying trip.'' \\
    & & \begin{minted}[escapeinside=||]{js}
{
  "onDate": ISODate("2016-04-30T17:10:29Z"),
  "onMode": 2,
  "onVid": 0,
  "onParentRoute": "",
  "onRouteId": 20,
  "onStopId": |{\bf 19936}|,
  "offDate": ISODate("2016-04-30T17:52:17Z"),
  "offMode": 2,
  "offVid": 0,
  "offParentRoute": "",
  "offRouteId": 1,
  "offStopId": 64408\
}, {
  "onDate": ISODate("2016-05-01T00:01:19Z"),
  "onMode": 2,
  "onVid": 0,
  "onParentRoute": "",
  "onRouteId": 1,
  "onStopId": 64408
}
    \end{minted}
    \\
    &  & Consistent.  Trip from Rosanna in afternoon; back from city so late that it's very early the following morning. Trains from Geelong do not seem to be in the dataset.  \\
          \hline
 \\
\pagebreak
 \hline
24 May 2016 & 8:59am & ``Our train driver advises that our Hursty Line service is on time and if we were travelling by car to the CBD it would be a 1hr 47min trip'' \\
   & & \begin{minted}[escapeinside=||]{js}
{
  "onDate": ISODate("2016-05-24T08:51:24Z"),
  "onMode": 2,
  "onVid": 0,
  "onParentRoute": "",
  "onRouteId": 20,
  "onStopId": 19933,
  "offDate": ISODate("2016-05-24T09:12:14Z"),
  "offMode": 2,
  "offVid": 0,
  "offParentRoute": "",
  "offRouteId": 18,
  "offStopId": 64407
}
   \end{minted}
   \\
   & & Consistent (8 mins).  On stop 19933 is Ivanhoe station, on the Hurstbridge line. \\
         \hline
 \\

21 Jun 2016 & 8:26am &  ``That moment at Rosanna when ur on wrong side of boom gates to catch the train...'' \\
   & & \begin{minted}[escapeinside=||]{js}
{
  "onDate": ISODate("2016-06-21T08:22:26Z"),
  "onMode": 2,
  "onVid": 0,
  "onParentRoute": "",
  "onRouteId": 20,
  "onStopId": |{\bf 19936 }|,
  "offDate": ISODate("2016-06-21T08:55:08Z"),
  "offMode": 2,
  "offVid": 0,
  "offParentRoute": "",
  "offRouteId": 18,
  "offStopId": 64407
}
\end{minted}
   \\
   & & Consistent. \\
         \hline
 \\
\pagebreak
 \hline
14 Sep 2016 & 6:51am & ``Bring on a new train station at Rosanna'' \\
   & & \begin{minted}[escapeinside=||]{js}
{
  "onDate": ISODate("2016-09-14T06:42:07Z"),
  "onMode": 2,
  "onVid": 0,
  "onParentRoute": "",
  "onRouteId": 20,
  "onStopId": |{\bf 19936}|,
  "offDate": ISODate("2016-09-14T07:26:50Z"),
  "offMode": 2,
  "offVid": 0,
  "offParentRoute": "",
  "offRouteId": 1,
  "offStopId": 64403\
}
   \end{minted}
\\
   & & Consistent (9 mins). \\
       \hline
 \\

2 Nov 2016 & 10:16pm &  ``Reading my copy of @beyondzeronews BZE Electric Vehicles report...on my electric train home.'' \\
   & & \begin{minted}[escapeinside=||]{js}
{
  "onDate": ISODate("2016-11-02T20:45:55Z"),
  "onMode": 2,
  "onVid": 0,
  "onParentRoute": "",
  "onRouteId": 1,
  "onStopId": 64403,
  "offDate": ISODate("2016-11-02T21:16:32Z"),
  "offMode": 2,
  "offVid": 0,
  "offParentRoute": "",
  "offRouteId": 20,
  "offStopId": |{\bf 19936}|
}
   \end{minted}
   \\
   & & Consistent.  \\
      \hline
 \\
\pagebreak
 \hline
20 Apr 2017 & 6:03pm & ``Waiting for city-bound train at Heidy so we can push thru to Rosanna...'' \\
   &  & \begin{minted}[escapeinside=||]{js}
{
  "onDate": ISODate("2017-04-20T17:45:52Z"),
  "onMode": 2,
  "onVid": 0,
  "onParentRoute": "",
  "onRouteId": 18,
  "onStopId": 19974,
  "offDate": ISODate("2017-04-20T17:59:38Z"),
  "offMode": 2,
  "offVid": 0,
  "offParentRoute": "",
  "offRouteId": 20,
  "offStopId": |{\bf 19936}|
}
   \end{minted}
   \\
   &  &  Touch-on at Clifton Hill rather than  Heidelberg, but this is consistent with changing at Heidelberg.  \\
       \hline
 \\

27 Apr 2017 & 8:03pm & ``I'll be checking out the buses home tonight. Back on the rails tomorrow from first train.'' \\
 & &  \begin{minted}[escapeinside=||]{js}
{
  "onDate": ISODate("2017-04-27T22:14:49Z"),
  "onMode": 2,
  "onVid": 0,
  "onParentRoute": "",
  "onRouteId": 1,
  "onStopId": 64403
}
\end{minted}
    \\
   & & This has a touch-on at a city train station, but no corresponding touch-off, which is consistent with a bus replacement. \\
      \hline
 \\
\pagebreak
 \hline
13 Dec 2017 & 8:28am & ``Today's Macleod-Heidelberg train replacement bus includes a cold bevvy and brekky.'' \\
    & & \begin{minted}[escapeinside=||]{js}
{
  "onDate": ISODate("2017-12-13T08:33:05Z"),
  "onMode": 2,
  "onVid": 0,
  "onParentRoute": "",
  "onRouteId": 20,
  "onStopId": 19983,
  "offDate": ISODate("2017-12-13T09:23:44Z"),
  "offMode": 2,
  "offVid": 0,
  "offParentRoute": "",
  "offRouteId": 1,
  "offStopId": 64403
}
    \end{minted}
    \\
    & & Consistent (5 mins).  Note the record is not for the bus but for the continuation from  Macleod station (19983).   \\
        \hline
 \\

6 Mar 2018 & 10:52pm & ``Heidy-Rosanna train replacement...to home. See you on same tomoz. That wil be a Rosanna-Heidy train replacement...to work.'' \\
    & & \begin{minted}[escapeinside=||]{js}
{
  "onDate": ISODate("2018-03-06T22:14:48Z"),
  "onMode": 2,
  "onVid": 0,
  "onParentRoute": "",
  "onRouteId": 1,
  "onStopId": 64403,
  "offDate": ISODate("2018-03-06T22:46:06Z"),
  "offMode": 2,
  "offVid": 0,
  "offParentRoute": "",
  "offRouteId": 20,
  "offStopId": 19935
}
     \end{minted}
     \\

    & & Tap-off at Heidelberg (19935), consistent with train replacement. Time matches (6 mins).\\
 \hline
 \\

\pagebreak
 \hline
27 Mar 2018 & 7:06am & ``Leave home 6am. Walk from Rosanna to Macleod station: 24 min. Express bus to Vic Park: 25 min. Train to Jolimont: 6 min. Time: 7.02am'' \\
    & & \begin{minted}[escapeinside=||]{js}
{
  "onDate": ISODate("2018-03-27T06:49:53Z"),
  "onMode": 2,
  "onVid": 0,
  "onParentRoute": "",
  "onRouteId": 18,
  "onStopId": 19975,
  "offDate": ISODate("2018-03-27T07:02:16Z"),
  "offMode": 2,
  "offVid": 0,
  "offParentRoute": "",
  "offRouteId": 18,
  "offStopId": 64407
}
     \end{minted}
     \\
    & & Perfectly matching time to the minute. 19975 is Victoria Park. \\
        \hline
 \\

\end{longtable}
\end{partialdetails}
\end{document}